\documentstyle{mn}

\input epsf.tex

\def\mag{^{\rm m}}
\def\Ic{{I_c}}
\def\Rc{{R_c}}
\def\rz{{r_z}}
\def\Vz{{V_z}}
\def\cI{{c_{\rm I_c}}}
\def\cV{{c_{\rm V}}}
\def\cR{{c_{\rm r}}}
\def\marc{{^{\rm m} /{\rm arcsec}^{2}}}
\def\mmu{{< \hspace{-3pt} \mu \hspace{-3pt}>}}
\def\mue{{{< \hspace{-3pt} \mu \hspace{-3pt}>}_{\rm e}}}
\def\Ie{{< \hspace{-3pt} I \hspace{-3pt}>_{\rm e}}}
\def\re{{r_{\rm e}}}
\def\mT{{m_{\rm T}}}
\def\MrT{{M_{\rm r_T}}}
\def\Ho50{{H_{\rm 0}=50~{\rm km\, s^{-1}\, Mpc^{-1}} }}

\title[The Evolution of cluster E and S0 galaxies: The Fundamental Plane]
{The evolution of cluster E and S0 galaxies measured from the Fundamental Plane\mbox{\Huge $^{\star}$} 
}
\author[I. J{\o}rgensen, M. Franx, J. Hjorth, P. G. van Dokkum]
{Inger J{\o}rgensen$^{1,2\,}$\mbox{\LARGE $^{\dag}$}\mbox{\LARGE $^{\ddag}$}\mbox{\LARGE $^{\S}$}
Marijn Franx$^{3,4,5\,}$\mbox{\LARGE $^{\ddag}$}\mbox{\LARGE $^{\S}$},
Jens Hjorth$^{6,7,8\,}$\mbox{\LARGE $^{\S}$}, 
Pieter G. van Dokkum$^{3,5\,}$\mbox{\LARGE $^{\S}$} \\
$^{1}$McDonald Observatory, The University of Texas at Austin, RLM 15.308, Austin, TX 78712, USA  \\
$^{2}$Gemini Observatory, 670 N.\ A`ohoku Pl., Hilo, HI 96720, USA (Postal address for IJ) \\
$^{3}$Kapteyn Institute, P.O.Box 800, 9700 AV Groningen, The Netherlands\\
$^{4}$Center for Astrophysics, 60 Garden Street, Cambridge, MA 02138, USA \\
$^{5}$Leiden Observatory, P.O.Box 9513, 2300 RA Leiden, The Netherlands (Postal address for MF and PvD)\\
$^{6}$Institute of Astronomy, Madingley Road, Cambridge CB3 0HA, UK \\
$^{7}$NORDITA, Blegdamsvej 17, DK-2100 Copenhagen {\O}, Denmark \\
$^{8}$Astronomical Observatory, University of Copenhagen, Juliane Maries Vej 30, DK-2100 Copenhagen {\O}, Denmark (Postal address for JH) }

\date{May 5, 1999, accepted for publication in Mon.\ Not.\ Royal Astron.\ Sco., Gemini Preprint \#43}

\begin{document}

\maketitle

\begin{abstract}
Photometry has been obtained for magnitude limited samples of 
galaxies in the two rich clusters Abell 665 (37 galaxies) and 
Abell 2218 (61 galaxies). 
Both clusters have a redshift of 0.18.
The limiting magnitude of the samples is $19\mag$ in the I-band.
Spectroscopy has been obtained for seven galaxies in A665 
and nine galaxies in A2218, all of which also have available photometry. 
Spectroscopy has been obtained for two
additional galaxies in A2218, one of which is a background galaxy.

Effective radii $\re$ and mean surface brightnesses $\Ie$ were 
derived from the photometry. 
The typical uncertainties are $\pm 0.078$ in $\log \re$ and
$\pm 0.12$ in $\log \Ie$. The combination $\log \re + 0.82 \log \Ie$
that enters the Fundamental Plane (FP) has an uncertainty of only
$\pm 0.018$.
The spectroscopy was used for measurements of the velocity dispersions,
$\sigma$.  The typical uncertainty is $\pm 0.023$ on $\log \sigma$.

The data are used to establish the FP, 
$\log \re = \alpha \log \sigma + \beta \log \Ie + \gamma$, 
for the clusters.
The FP for these two clusters adds important
knowledge about the properties of E and S0 galaxies in the relatively
unexplored redshift interval 0.05 to 0.3.
We have compared the FP for A665 and A2218 with the FP for the three
clusters CL0024+16, CL1358+62 and MS2053-04 with redshifts between
0.33 and 0.58, and with the FP for the Coma cluster.
The scatter around the FP is similar for all six clusters.
We find that the FP for the intermediate redshift clusters has a 
smaller coefficient $\alpha$ than found for the Coma cluster
and other nearby clusters.
This may either be caused by selection effects for the 
intermediate redshift clusters or by differences in the evolution of
low luminosity galaxies and high luminosity galaxies.

The mass-to-light (M/L) ratios, as measured from the FP, change with 
redshift. At $z$=0.18 the M/L ratio for photometry in Gunn $r$ in the 
rest frame of the clusters is \mbox{($11\pm 7$)\%} smaller than for 
the Coma cluster.
Using the data for A665 and A2218
together with the previously published data for CL0024+16, CL1358+62 
and MS2053-04, we find that the M/L ratios for photometry calibrated
to Gunn $r$ change with redshift as
$\Delta \log M/L_r = (-0.26\pm 0.06) \Delta z$ for $q_{\rm o} = 0.5$.
This change in the M/L ratio is equivalent to the absolute magnitudes 
changing as $\Delta \MrT = (-0.65\pm 0.15)\Delta z$.
These new results are consistent with the previously published
analysis for CL0024+16, CL1358+62 and MS2053-04.

For $q_{\rm o} =0.5$ the results are consistent with passive evolution
of a stellar population which formed at a redshift larger than five.
For $q_{\rm o} =0.15$ the formation redshift must be larger than 1.7.
Our new data for A665 and A2218 confirm the gradual and slow 
evolution of the bright E and S0 galaxies.
However, possible star formation in E and S0 galaxies during the 
last 3-4 Gyr of the history of the Universe, 
as well as the selection effects complicate the interpretation of 
the data.

\end{abstract}

\begin{keywords}
galaxies: clusters: individual: Abell 665, Abell 2218 --
galaxies: elliptical and lenticular, cD --
galaxies: evolution -- 
galaxies: stellar content.
\vspace*{-1.9cm}
\end{keywords}

\vspace*{1.9cm}
\section{Introduction}

\footnotetext[1]{
Based in part on observations made with the NASA/ESA 
{\sl Hubble Space Telescope} obtained at the Space Telescope Science 
Institute, which is operated by AURA under NASA contract NAS5-26555;
the Multiple Mirror Telescope, a joint 
facility of the Smithsonian Institution and the University of Arizona;
and the Nordic Optical Telescope,
operated on the island of La Palma jointly by Denmark, Finland,
Iceland, Norway and Sweden, in the Spanish Observatorio del 
Roque de los Muchachos of the Instituto de Astrofisica de Canarias.}

\footnotetext[2]{Hubble Fellow}

\footnotetext[3]{
Visiting Astronomer, Kitt Peak Observatory, a division of National 
Optical Astronomy Observatories, which is operated by AURA under 
contract with the National Science Foundation.}
\footnotetext[4]{
E-mail: ijorgensen@gemini.edu, franx@strw.leidenuniv.nl, \newline
jens@astro.ku.dk, dokkum@strw.leidenuniv.nl}

Observational studies show that the formation and evolution of
galaxies is a complex process, which may involve 
interactions, star bursts and infall (e.g., 
Dressler et al.\ 1994ab; Lilly et al.\ 1996; Moore et al.\ 1996).
It has also been found that some nearby E and S0 galaxies may have
experienced star formation in the last 3-4 Gyr. 
Caldwell et al.\ (1993) found
some E and S0 galaxies in the Coma cluster to have post-starburst
spectra, and Faber et al. (1995) suggest that nearby field E 
galaxies have a substantial variation in the mean age of their
stellar populations.

In order to study the evolution of galaxies we need to study both the 
morphological evolution as well as the evolution of the luminosities and
the mass-to-light (M/L) ratios of the galaxies.
High-resolution imaging with the {\sl Hubble Space Telescope} (HST)
and from the ground, combined with spectroscopy from the ground
make it possible to carry out this kind of studies.
Studies of the morphological evolution show that the fraction of
spiral galaxies and irregular galaxies in clusters was higher
at larger redshifts (e.g., Dressler et al.\ 1994ab; Oemler et al.\ 1997;
Dressler et al.\ 1997; Couch et al.\ 1998).
A detailed investigation of how the luminosities  of disk galaxies
change with redshift is  possible using the Tully-Fisher (1977) 
relation. Vogt et al.\ (1996) established the Tully-Fisher relation for
a sample of field galaxies with redshifts between 0.1 and 1 and
found a luminosity evolution in the B-band of $\approx 0\fm 6$
between $z=1$ and the present.

The Fundamental Plane (FP) (Dressler et al.\ 1987;
Djorgovski \& Davis 1987) for elliptical galaxies makes it possible
to study how the luminosities and the M/L ratios of these galaxies
change with redshift.
Also, S0 galaxies in nearby clusters follow the FP (e.g., J{\o}rgensen,
Franx \& Kj\ae rgaard 1996, hereafter JFK96).
The FP relates the effective radius, $\re$, the mean surface brightness
within this radius, $\Ie$, and the (central) velocity dispersion, 
$\sigma$, in a tight relation, which is linear in log-space.
For 226 E and S0 galaxies in nearby clusters JFK96 found
$\log \re = 1.24 \log \sigma - 0.82 \log \Ie + {\rm cst}$,
for photometry in Gunn $r$.
This relation may be interpreted as $M/L \propto M^{0.24} \re^{-0.02}$ 
(cf., Faber et al.\ 1987).
The change of the M/L ratio with the mass is partly caused by
changes in the stellar populations.
Some of the apparent change may also be explained by
non-homology of the galaxies (e.g., Hjorth \& Madsen 1995).
Throughout this paper we assume that the FP implies a relation 
between the M/L ratios and the masses of the galaxies. 

The scatter of the FP is very low, equivalent to a scatter of
23\% in the M/L ratio (e.g., JFK96).
Thus, the FP offers the possibility of detecting even small
differences in the M/L ratios by observing a handful of galaxies
in a distant cluster.

The FP has been used to study the evolution of the M/L ratios of 
cluster galaxies as a function of redshift up to a redshift of 0.83
(van Dokkum \& Franx 1996; Kelson et al.\ 1997; Bender et al.\ 1998;
van Dokkum et al.\ 1998; Pahre, Djorgovski \& de Carvalho 1999). 
All of these studies find the M/L ratios of the galaxies to increase
slowly with decreasing redshift.
Only the study by Pahre et al.\ contains clusters with redshifts between
0.05 and 0.3.
In this paper we establish the FP for the two rich clusters
Abell 665 and Abell 2218. Both clusters have a redshift of 0.18,
adding data to a redshift interval not well-covered by the previous
studies.

We use the data for A665 and A2218, data for the three clusters studied
by van Dokkum \& Franx (1996) and Kelson et al.\ (1997),
and data for the Coma and the HydraI clusters 
to study how the FP and the M/L ratios 
of the E and S0 galaxies change with redshift.
The full sample covers about half the age of the Universe.

\begin{table*}
\begin{minipage}{13.2cm}
\caption{Instrumentation \label{tab-inst} }
\begin{tabular}{lrrrr}
   & \multicolumn{2}{c}{Photometry} & \multicolumn{2}{c}{Spectroscopy} \\ \hline
Dates     &  March 4-9, 1994 & Sept.-Dec., 1994 & Jan.\ 18-19, 1991 & June 6-8, 1994 \\
Telescope & NOT, LaPalma     & HST & MMT           & KPNO 4m        \\
Camera/Spec. & Astromed   & WFPC2 & Red Channel Spec. & R\&C Spec. \\
CCD       & EEV 1152$\times$770 & Loral & TI5           & T2KB           \\
r.o.n.    & 4.5 e$^-$        & 5.0 e$^-$ & 9.0 e$^-$ & 7.4 e$^-$ \\
gain      & 0.9 e$^-$/ADU    & 7.5 e$^-$/ADU & 2.0 e$^-$/ADU & 2.6 e$^-$/ADU \\
Pixel scale & 0.1635\arcsec/pixel& 0.0996\arcsec/pixel$^a$ & 0.6\arcsec/pixel  & 0.69\arcsec/pixel \\
Slit width  & ...            & ... & 1.8\arcsec    & 1.9\arcsec \\
Spectral res., $\sigma$& ... & ... & 1.71\AA       & 1.38\AA \\
Wavelength range & ...       & ... & 5780-6420\AA  & 4255-6970\AA$^b$ \\
Cluster(s)  & A665, A2218 & A665, A2218 & A665 & A2218 \\ \hline
\end{tabular}

Notes. $^a$ All galaxies studied in this paper are imaged on the WF
chips of WFPC2. $^b$ The unvignetted range is 4550-6500\AA.
\end{minipage}
\end{table*}

The available photometry and spectroscopy, as well as the
derivation of the photometric and spectroscopic parameters
is described in Sections 2 and 3, respectively.
The details of the data reduction are provided in Appendix A and B.
Appendix A also contains photometric parameters (effective radii,
mean surface brightnesses and colors) of magnitude limited
samples of galaxies in A665 and A2218.
Section 4 briefly describes the data for the other clusters.
The FP is analyzed in Section 5.
In this section we also discuss the evolution of the M/L ratios 
of the galaxies.
The conclusions are summarized in Section 6.

\section{The photometry}

The central parts of the two rich clusters Abell 665 and Abell 2218
were observed with the Nordic Optical Telescope (NOT), La Palma,
in March 1994.
Observations were done in the I-band and the V-band.
The sizes of the observed fields are     
$3\farcm 4 \times 2\farcm 7$ for A665 and $2\farcm 8 \times 3\farcm 5$
for A2218.
\mbox{Table \ref{tab-inst}} summarizes the instrumentation for the 
observations.
The I-band images were taken as individual 600 sec 
exposures, with a few shorter exposures.
The total exposure time, the seeing and the sky brightness of
each of the final co-added images are given in Table \ref{tab-NOT}.
The V-band images were taken on another night than the 
I-band images, and have rather poor seeing.

\begin{table}
\caption{NOT images \label{tab-NOT} }
\begin{tabular}{lcrrr}
Image & Filter & $t_{\rm exp}$ & seeing & $m_{\rm sky}$ \\
      &        & [sec]         & [$''$] & [$\marc$]     \\ \hline
A665  field 1 & I & 9600 & 0.66 & 20.0 \\
A665  field 1 & V & 1200 & 1.86 & 21.8 \\
A665  field 2 & I & 5874 & 0.60 & 19.6 \\
A665  field 2 & V & 1200 & 2.03 & 21.8 \\
A2218 field 1 & I & 5400 & 0.70 & 19.3 \\
A2218 field 1 & V & 1200 & 3.33 & 21.2 \\
A2218 field 2 & I & 5400 & 0.82 & 19.4 \\
A2218 field 2 & V & 1200 & 4.23 & 21.2 \\ \hline
\end{tabular}

Notes. The seeing is measured as the full-width-half-maximum (FWHM) of
a Gaussian fit to the core of images of stars. 
A665 field 1 is centered 0\farcm6 east and 0\farcm 3 south of the BCG;
field 2 is offset 1\farcm7 to the west of field 1.
A2218 field 1 is centered 0\farcm3 to the south of the BCG;
field 2 is offset 1\farcm7 to the north of field 1.
\end{table}

\begin{table}
\caption{HST images \label{tab-HST} }
\begin{tabular}{lrr}
Image        & Filter & $t_{\rm exp}$ [sec] \\ \hline
A665 field 1 & F606W  & 4400 \\
A665 field 1 & F814W  & 4400 \\
A665 field 2 & F606W  & 5100 \\
A665 field 2 & F814W  & 4800 \\
A2218        & F702W  & 6500 \\ \hline
\end{tabular}

Notes.
A665 field 1 is centered 0\farcm3 south-east of the BCG.
A665 field 2 is centered 1\farcm7 west of the BCG.
The observation of A2218 is centered 0\farcm9 east of the BCG.
\end{table}

A665 was observed with the HST with the 
Wide-Field-Planetary-Camera 2 (WFPC2) in the filters F606W and F814W.
Two fields were observed giving a total coverage of approximately 
9.4 square arc-minutes on the WF chips.
For A2218 we use HST/WFPC2 archival data obtained in the F702W 
filter.  One field in Abell 2218 was observed with the HST, covering
approximately 4.7 square arc-minutes on the WF chips.
The images were taken as individual exposures with exposure
times between 1600 sec and 2300 sec. The total exposure times
for the co-added images are given in \mbox{Table \ref{tab-HST}.}

The basic reductions of the images are described in Appendix A, which
also covers the calibration of the photometry to the standard passbands
$\Ic$ and $V$.
In the following we describe the determination of the global 
photometric parameters (effective radii and mean surface brightnesses)
and the calibration of the photometry to Gunn $r$ and Johnson V in
the rest frames of the clusters.

\subsection{Global parameters}

Determination of the effective radius, $\re$, and the mean surface 
brightness inside this radius, $\mue$, was done following the technique
described by van Dokkum \& Franx (1996).
This technique uses the full 2-dimensional information in the image.
Each galaxy is fit with a 2-dimensional model,
taking the point-spread-function (PSF) into account.
The output from the fit is the center of the galaxy ($x$,$y$),
the effective radius, $\re$, the surface brightness at the effective
radius, $I_{\rm e}$, the ellipticity, $\epsilon$, the position angle
of the major axis and the level of the sky background.
All galaxies presented here were fitted with $r^{1/4}$ profiles.
The mean surface brightness $\mue$ was derived from the local
surface brightness at $\re$ under the assumption that the galaxies
have $r^{1/4}$ profiles.

The PSF images for the NOT I-band images were constructed from 2-3 
well-exposed stars in each image.
The PSF images have a size of $10\farcs 5 \times 10\farcs 5$.
It was tested that the PSFs did not vary over the field.
The PSF images for the HST observations were constructed using 
Tiny Tim (Krist \& Hook 1997) and have a size of $3''\times 3''$.

All galaxies brighter than an apparent total magnitude in the I-band of
$\mT \approx 20\mag$ were fitted.
The brightest cluster galaxies (BCG) in both clusters were 
fitted within a radius of 1-2$\re$.
The fainter galaxies were fitted within the radius where 
the local surface brightness reached $\approx 26\marc$ in the I-band.
This radius is typically 5-7$\re$.

Before fitting the fainter galaxies the BCGs
were modeled with the program {\sc Galphot} 
(Franx, Illingworth \& Heckman 1989a; 
J{\o}rgensen, Franx \& Kj{\ae}rgaard 1992).
This program fits a model of ellipses to the isophotes of the galaxy.
The model was then subtracted.
The use of {\sc Galphot} ensures that gradients in the background
due to signal from the BCG are properly
removed before fitting the fainter galaxies.
Further, the isophotes for the BCG in A2218 are not co-centric,
as also noticed by Kneib et al.\ (1995).
{\sc Galphot} handles this properly while the fitting technique 
described above uses a fixed center and therefore leaves large 
residuals, which in turn affect the results for the fainter galaxies.
The BCGs were also modeled with the 2-dimensional fitting technique.
The effective radii derived from this agree within 0.02 in $\log \re$ 
with the results from fits to the growth curves produced by 
{\sc Galphot}.
Fitting the growth curves produced by {\sc Galphot} does not give
the possibility of 2-dimensional modeling the effect of the PSF.
In the following, we therefore use results from the 2-dimensional
fitting technique.

\begin{table*}
\begin{minipage}{17.7cm}
\caption[]{Photometric parameters \label{tab-phot} }
\begin{tabular}{lrrrrrrrrrrrrr}
       & \multicolumn{7}{c}{\rule[1mm]{33mm}{0.2mm} NOT \rule[1mm]{33mm}{0.2mm} } & 
\multicolumn{6}{c}{\rule[1mm]{31mm}{0.2mm} HST \rule[1mm]{31mm}{0.2mm} } \\
\multicolumn{2}{l}{Galaxy \hfill Field} & $\log \re$ & $\mue$ & $\frac{\chi^2}{N_{\rm pix}}$ & $\mue$ & $\mue$ & ($V-\Ic $) &
         $\log \re$ & $\mue$ & $\frac{\chi^2}{N_{\rm pix}}$ & $\mue$ & $\mue$ & ($V-\Ic $) \\ 
       & & {\small [arcsec]} & ($\Ic$) & & ($\rz $) & ($V_z$) &
       & {\small [arcsec]} & ($\Ic$) & & ($\rz $) & ($V_z$) \\ \hline
A665-1150$^a$ & 1 & 1.012 & 21.82 & 2.9 & 22.67 & 22.96 & 1.54 & 1.056 & 21.97 &  5.2 & 22.82 & 23.11 & 1.53 \\
A665-1168     & 1 & 0.333 & 19.78 & 2.1 & 20.63 & 20.92 & 1.54 & 0.376 & 19.93 &  1.3 & 20.78 & 21.06 & 1.52 \\
A665-1196     & 1 &-0.352 & 17.73 & 9.0 & 18.58 & 18.86 & 1.51 &-0.149 & 18.58 &  7.9 & 19.43 & 19.68 & 1.47 \\
A665-2105$^b$ & 2 & 0.876 & 22.09 &29.7 & 22.92 & 23.07 & 1.20 & 1.003 & 22.42 & 18.5 & 23.23 & 23.41 & 1.24 \\
A665-2108     & 2 &-0.205 & 19.48 & 1.5 & 20.33 & 20.61 & 1.53 & ... & ... & ... & ... & ...  \\
A665-2111     & 2 &-0.170 & 18.56 & 6.7 & 19.41 & 19.74 & 1.62 &-0.011 & 19.17 &  4.1 & 20.02 & 20.29 & 1.51 \\
A665-2144     & 2 &-0.189 & 18.83 & 1.6 & 19.68 & 19.96 & 1.52 &-0.096 & 19.22 &  1.6 & 20.06 & 20.29 & 1.41 \\ 
A2218-L118    & 1 & 0.051 & 19.20 & 5.5 & 20.06 & 20.39 & 1.62 & 0.097 & 19.30 &  3.1 & 20.16 & 20.48 & ... \\
A2218-L148    & 1 &-0.075 & 18.53 & 3.1 & 19.39 & 19.70 & 1.59 & 0.067 & 19.03 &  6.4 & 19.89 & 20.20 & ... \\
A2218-L198    & 1 &-0.061 & 19.14 & 2.0 & 19.99 & 20.29 & 1.55 &-0.143 & 18.81 &  1.9 & 19.67 & 19.96 & ... \\
A2218-L244    & 1 & 0.934 & 22.01 & 2.0 & 22.87 & 23.15 & 1.53 & 0.930 & 22.02 &  1.4 & 22.88 & 23.16 & ... \\
A2218-L341    & 1 &-0.124 & 18.27 & 2.6 & 19.13 & 19.42 & 1.56 &-0.116 & 18.26 &  4.1 & 19.11 & 19.41 & ... \\
A2218-L391$^a$& 1 & 1.403 & 23.07 & 3.6 & 23.92 & 24.18 & 1.48 & 1.412 & 23.25 & 21.0 & 24.10 & 24.37 & ... \\
A2218-L430    & 2 &-0.033 & 19.06 & 1.4 & 19.92 & 20.24 & 1.62 &-0.061 & 18.95 &  1.0 & 19.81 & 20.13 & ... \\
A2218-L482    & 2 & 0.036 & 18.77 & 2.2 & 19.63 & 19.95 & 1.62 & ... & ... & ... & ... & ... \\
A2218-L535    & 2 & 0.238 & 19.63 & 1.9 & 20.49 & 20.82 & 1.63 & ... & ... & ... & ... & ... \\ \hline
\end{tabular}

Notes.
$^a$ Brightest cluster galaxy.  $^b$ E+A galaxy. 
Identifications of galaxies in A2218 are from Le Borgne et al.\ (1992).
Field -- the NOT field that contains the galaxye, see Table \ref{tab-NOT} 
for seeing values.
The observed magnitudes labeled $\Ic$ and $V$ and the colors $(V-\Ic )$
are standard calibrated to the Johnson-Kron-Cousins photometric system,
and corrected for galactic extinction.
The formal fitting uncertainties are typically as follows.
$\log \re$: $\pm 0.006$ for NOT data, $\pm 0.002$ for HST data.
$\mue$: $\pm 0.02$ for NOT data, $\pm 0.007$ for HST data.
The typical (random) uncertainty on the $(V-I_c)$ is 0.01.
The goodness-of-fit is given as $\chi^2/N_{\rm pix}$, 
where $N_{\rm pix}$ is the number of pixels within the fitting radius.
$\chi^2 = \sum (n_i-n_{i,\rm model})^2/\sigma_i^2$, 
with the sum over all the pixels within the fitting radius. 
$n_i$ is the signal in pixel $i$,
$n_{i,\rm model}$ is the model value in the pixel, and $\sigma_i^2$ the
noise (photon- and read-out-noise) in the pixel.

Magnitudes $\rz$ are fully corrected Gunn $r$ 
magnitudes in the rest frames of the clusters, see equations 
(\ref{eq-A665GR}) and (\ref{eq-A2218GR}).
Magnitudes $\Vz$ are fully corrected Johnson V
magnitudes in the rest frames of the clusters, see equations
(\ref{eq-A665V}) and (\ref{eq-A2218V}).
\end{minipage}
\end{table*}

Table \ref{tab-phot} lists the derived effective parameters.
Only galaxies for which we also have spectroscopy are presented in this
table. Tables \ref{tab-photA665} and \ref{tab-photA2218} contain the 
derived effective parameters for all galaxies brighter than $19\mag$ 
and within the fields covered by the NOT images.
The magnitudes were standard calibrated as described in Appendix A.
The $\Ic$ magnitudes from the HST data have been offset to consistency 
with the ground-based magnitudes, see Appendix A.
Table \ref{tab-phot} gives typical values of the formal fitting
uncertainties. However, a comparison of the NOT and the HST data
results in more realistic estimates of the uncertainties, see
Appendix A.
We find the uncertainties of $\log \re$ and $\mue$ to be 
$\pm 0.078$ and $\pm 0.29$, respectively.
Because of the correlation of the errors in $\log \re$ and $\mue$,
the combination $\log \re - 0.328 \mue$ that enters the FP has an
uncertainty of only $\pm 0.018$.
In the analysis of the FP for the clusters (Section 5), we use
the average of the effective parameters determined from the HST 
observations and the NOT observations, whenever results from both
telescopes are available.

\subsection{Colors}

The colors of the galaxies were determined from the ground-based
observations. First, the I-band images were convolved to the seeing of 
the V-band images. 
This minimizes the errors in the derived colors due to mismatched
seeing in the two passbands.
The colors were derived within apertures with radii 
1\farcs64 (10 pixels) and 2\farcs94 (18 pixels) for A665 and A2218, 
respectively.
The larger aperture for A2218 was necessary due to the very poor seeing
of the V-band images.
The colors within these apertures are in the following used as global 
colors of the galaxies, see Table \ref{tab-phot} for the sample that
also has spectroscopy and Tables \ref{tab-photA665} and 
\ref{tab-photA2218} for all galaxies brighter than $19\mag$.
For nearby E and S0 galaxies, the color gradients in the local surface 
brightnesses are usually of the order $0\fm 05$
per dex in the radius, for colors based on Johnson B and Gunn $r$
(e.g., J{\o}rgensen et al.\ 1995a).
The color gradients in the aperture magnitudes are smaller, typically
less than $0\fm 04$ per dex in the radius.
Assuming that the color gradients of the galaxies at redshift 0.18
are similar to those of nearby galaxies,
the derived colors are expected to deviate less than
$\pm 0.02$ from global colors within the effective radii.
This results in a negligible effect of less than $0\fm 002$ on the
calibrated magnitude in Gunn $r$ in the rest frame of the clusters,
cf.\ Sect.\ \ref{sec-restframe}.

For the galaxies in A665 colors were also derived from the HST images.
We used apertures with radii of 1\farcs5.
The aperture size is smaller than those used for the NOT images
in order to take advantage of the better spatial resolution of the
HST images and limit the amount of contamination from neighboring
objects.
No correction for zero point offsets between the ground-based
photometry and the HST photometry were applied to the colors,
since we have no reliable way of checking the zero point for the
HST V magnitudes. Thus, we implicitly assume that the zero point
difference between the ground-based photometry and the HST photometry
is the same in the I-band and the V-band.
The colors derived from the HST observations are on average
0.04 mag bluer than the ground-based colors, cf.\ Table \ref{tab-phot}. 
This may be caused by errors in the HST colors due to the uncertainties
in the magnitude zero points. Alternatively, uncertainties may be
introduced by the poor seeing of the ground-based V images.
The effect on the magnitudes calibrated to \mbox{Gunn $r$} in the
rest frame of the cluster will be less than $0\fm 003$,
cf.\ Sect.\ \ref{sec-restframe}.

\begin{table*}
\begin{minipage}{14.0cm}
\caption[]{Uncertainties affecting the photometric calibration \label{tab-unphot} }
\begin{tabular}{lllrr}
Cluster & Tel.\ & Calibration/Parameter & Random unc.\  & Systematic unc. \\ \hline
A665    & NOT   & Instrumental $\Ic$ &        & 0.01 \\
A2218   & NOT   & Instrumental $\Ic$ &        & 0.015 \\
A665,A2218 & NOT & Standard $\Ic$    & 0.021  & \\
A665,A2218 & NOT & Standard $V$      & 0.014  & \\
A665,A2218 & NOT & Standard $(V-\Ic)$& 0.025  & \\
A665    & HST   & Holtzman et al.\ (1995) $\Ic$  & 0.015 & \\
A665    & HST   & Holtzman et al.\ (1995) $(V-\Ic)$ & 0.028 & \\
A665    & HST   & Offset to ground-based $\Ic$ & & 0.009 \\
A2218   & HST   & Holtzman et al.\ (1995) $\Ic$  & 0.017$^a$ & \\
A2218   & HST   & Offset to ground-based $\Ic$ & & 0.013 \\
A665,A2218 & NOT,HST & $(V-\Ic)_{\rm meas}$ vs.\ $(V-\Ic)_{\re}$ & 0.02 & \\
A665,A2218 & NOT & Total $(V-\Ic)$  & 0.032 & \\
A665    & HST & Total $(V-\Ic)$  & 0.034 & \\
A665,A2218 & NOT,HST & Gunn $r$ in rest frame, transformation & 0.02$^b$ & \\
A665,A2218 & NOT,HST & Johnson V in rest frame, transformation & 0.02$^b$ & \\ \hline
A665  & NOT & Total $\rz$ & 0.028  & 0.01 \\
A665  & HST & Total $\rz$ & 0.024  & 0.02 \\
A665  & NOT & Total $\Vz$ & 0.032  & 0.01 \\
A665  & HST & Total $\Vz$ & 0.029  & 0.02 \\
A2218 & NOT & Total $\rz$ & 0.028  & 0.015 \\
A2218 & HST & Total $\rz$ & 0.024  & 0.028 \\
A2218 & NOT & Total $\Vz$ & 0.032  & 0.015 \\
A2218 & HST & Total $\Vz$ & 0.029  & 0.028 \\ \hline
\end{tabular}

Notes. $^a$ Contributions from the transformation $(V-\Rc ) = 0.52 (V-\Ic )$ and 
the transformation from Holtzman et al.\ (1995).
$^b$ Contribution from transformation, only.

\end{minipage}
\end{table*}

\subsection{Calibration to the rest frame of the clusters 
\label{sec-restframe} }

In order to compare the FP for A665 and A2218 with results for
nearby clusters we calibrate the photometry to \mbox{Gunn $r$} in the 
rest frame of the clusters.  Following Frei \& Gunn (1994),
the magnitude in a broad passband can be written as
$m_i = c_i - 2.5 \log F(\nu _i)$, with $c_i \equiv \Delta b_i - 48.60$.
$\Delta b_i$ are given by Frei \& Gunn.
$F(\nu _i)$ is the flux density at the effective wavelength
of the passband.
We assume the flux density in the redshifted \mbox{Gunn $r$} band is 
related to the observed V-band and $\Ic$-band as
$F(\nu _r (z) ) = F(\nu _V) ^{\alpha} F(\nu _{\Ic}) ^{1-\alpha}$,
cf.\ van Dokkum \& Franx (1996).
The magnitudes in Gunn $r$ in the rest frame of the clusters can
then be written
\begin{equation}
\label{eq-GR}
\rz = \Ic + \alpha (V-\Ic ) - \cI - \alpha (\cV - \cI ) + \cR  + 2.5 \log (1+z)
\end{equation}
where $\alpha$ is derived from the spectral energy distribution.
The $\Ic$ magnitudes and the colors $(V-\Ic )$ are the observed standard
calibrated values, corrected for galactic extinction.

A665 has z=0.181 (Oegerle et al.\ 1991).
From the spectral energy distributions for E and Sa galaxies given by
Coleman, Wu \& Weedman (1982) we find $\alpha _{\rm A665} = 0.053$.
The  resulting calibration to reach Gunn $r$ in the rest frame is
\begin{equation}
\label{eq-A665GR}
\rz = \Ic + 0.053 (V-\Ic ) + 0.77
\end{equation}
A2218 has z=0.176 (Le Borgne, Pell\'{o} \& Sanahuja 1992).
We find $\alpha _{\rm A2218} =0.062$, and the resulting calibration is
\begin{equation}
\label{eq-A2218GR}
\rz = \Ic + 0.062 (V-\Ic ) + 0.76
\end{equation}
The mean surface brightnesses, $\mue$, calibrated to Gunn $r$ in the 
rest frame of the clusters are given in Table \ref{tab-phot}.

Table \ref{tab-phot} also lists the mean surface brightnesses 
calibrated to Johnson V in the rest frame of the clusters.
The calibrations to reach Jonhson V are 
\begin{equation}
\label{eq-A665V}
V_z = \Ic + 0.48 (V-\Ic ) + 0.40
\end{equation}
for A665 and
\begin{equation}
\label{eq-A2218V}
V_z = \Ic + 0.49 (V-\Ic ) + 0.39
\end{equation}
for A2218.

Because the photometry for the best comparison sample 
(the Coma cluster) is available in Gunn $r$,
and also because of the uncertainty in the derived colors, cf.\ Section
2.2, we will primarily use the mean surface brightnesses calibrated to 
Gunn $r$ throughout the rest of this paper.

\subsection{Summary of uncertainties}

Table \ref{tab-unphot} summarizes the different sources on uncertainty
affecting the photometric calibration.
The entries noted as ``Total'' uncertainty give the total uncertainty on 
the listed parameter, while all other entries give the contribution from
the individual calibration step.
The uncertainties on the instrumental $\Ic$ magnitudes for the NOT data
refer to the consistency of the magnitude zero points of the co-added 
images, cf.\ Sect.\ A1.
The uncertainties on the standard calibrated magnitudes for the NOT data
are the rms scatter of the standard transformations, cf.\ Sect.\ A3.
For the HST data the uncertainties consist of random uncertainties due
to the uncertainties in the transformations adopted from Holtzman et al.\
(1995), cf.\ Sect.\ A3, and systematic uncertainties in matching the
HST photometry to the ground-based system.
The colors $(V-\Ic)$ are subject to an additional uncertainty because 
we measure the colors within fixed aperture sizes rather than within 
the effective radii, cf.\ Sect.\ 2.2. 
Finally, the transformations used to calibrate the magnitudes to 
Gunn $r$ and Johnson $V$ in the rest frames of the clusters contribute
to the uncertainties.
The last part of Table \ref{tab-unphot} summarizes the total uncertainties,
random or systematic, on the calibrated magnitudes $\rz$ and $\Vz$.
The magnitudes have typical random uncertainties of $0\fm 03$, and typical
systematic uncertainties of $0\fm 02$.

\section{The spectroscopy}

A665 galaxies were selected for the spectroscopic observations
from a ground-based R-band image kindly provided by
M.\ Birkinshaw. The flux of the galaxies within a $1\farcs 8\times 8''$
aperture was used as selection criterion.
Spectra of 7 galaxies in A665 were obtained with the MMT in
January 1991, cf.\ Table \ref{tab-inst}.
The total integration time was 9 hours.
For one of the galaxies the signal-to-noise was too low to derive
a reliable value of the velocity dispersion.

For A2218 we used the catalog by Le Borgne, Pell\'{o} \& Sanahuja 
(1992) as the main source for the selection of galaxies, 
supplemented with the catalog by Butcher, Oemler \& Wells (1983).
Galaxies were selected on magnitude and color. The $(B-R)$ color was
required to be within 0.15 of the color-magnitude relation for the 
red cluster galaxies. 
One redder galaxy, BOW97, was included.
This galaxy turned out to be a background galaxy.
Spectra of 10 galaxies in A2218 were obtained with the KPNO 4m
in June 1994, cf.\ Table \ref{tab-inst}.
The total integration time was 11.8 hours.

The basic reductions of the spectroscopy are described in
Appendix B, which  also shows the final spectra of the galaxies.
Here we describe the determination of the velocity dispersions
(see also Franx 1993ab for the galaxies in A665).

\subsection{Instrumental resolution and template stars}

Determination of velocity dispersions of galaxies is usually
done with spectra of template stars obtained with the same 
instrumentation as the galaxy spectra.
Due to the large redshifts of A665 and A2218 this will not work
for these clusters.
A further complication is that the spectra of A665 and A2218 were
obtained through multi-slit masks and the instrumental resolution 
varies with wavelength and from slit-let to slit-let.

\begin{figure}
\epsfxsize=8.8cm
\epsfbox{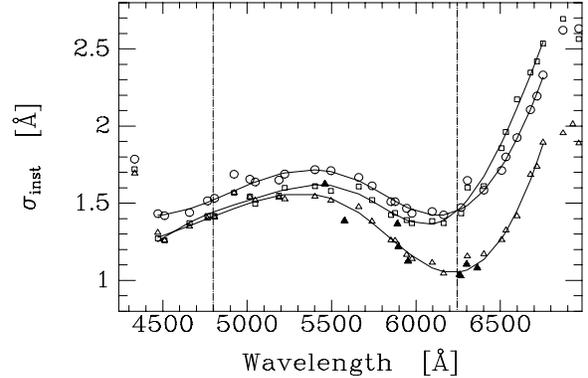}
\caption[ ]{The instrumental resolution for the A2218 spectra,
as derived from He-Ne-Ar calibration spectra.
Triangles -- central slit-let; boxes and circles -- the two slit-lets
closest to the edges of the CCD.
The filled symbols are based on sky lines.
The resolution derived from the sky lines agree with the resolution
derived from the calibration spectra.
The solid lines show the fitted functions.
The dot-dashed lines show the approximate wavelength interval,
which was used for determination of the velocity dispersions.
\label{fig-resolution} }
\end{figure}

Following van Dokkum \& Franx (1996),
we mapped the instrumental resolution from He-Ne-Ar lamp exposures.
Further, the instrumental resolution was determined from the
sky lines in the final average spectra of the galaxies.
In all cases Gaussian profiles were fitted to the emission lines.
Fig.\ \ref{fig-resolution} shows the instrumental resolution
versus wavelength for representative slit-lets for the A2218 spectra.
The determinations from lamp exposures and from sky lines agree.
Low order polynomials were fitted to the instrumental resolution
versus wavelength for each slit-let.

We used a spectrum of HD192126 as the template star for the A665 
spectra.
The spectrum of the star was obtained with the Canada-France-Hawaii 
Telescope by K.\ Kuijken. The wavelength range for the spectrum is
4650-5650\AA, with a resolution of $\sigma _{\rm inst}=0.56$\AA.
As template stars for the A2218 spectra we used
spectra of HD56224, HD72324 and HD102494
obtained with the Keck Telescope in May 1996 by D.\ Kelson.
The wavelength range of these spectra is 4002-5314\AA, with
a resolution of $\sigma _{\rm inst}=0.9$\AA.

For each galaxy spectrum the spectra of the template stars were
convolved to the instrumental resolution of the galaxy spectrum.
The variation of the resolution with wavelength was taken into account.

\subsection{Fourier Fitting}

The velocity dispersions, $\sigma$, were determined by fitting the 
galaxy spectra with the template spectra using the Fourier Fitting 
Method (Franx, Illingworth \& Heckman 1989b).
The Fourier transforms of the spectra were filtered to avoid
the low frequencies.
We used a limit equivalent to $\approx$(100\AA )$^{-1}$ in the rest
frame of the clusters. 
This is the same limit used by 
J{\o}rgensen, Franx \& Kj\ae rgaard (1995b).
In the case of A2218,
the final velocity dispersion is the average of the determinations
with the three different template stars.
The formal fitting uncertainty on $\log \sigma$ is typically 0.023.
For A665 the mean spectra over the central 1\farcs 8 were fitted.
Thus, the aperture size is 1\farcs 8$\times$1\farcs 8.
The aperture size used for A2218 was 1\farcs 9$\times$3\farcs 45.

\begin{figure}
\epsfxsize=8.8cm
\epsfbox{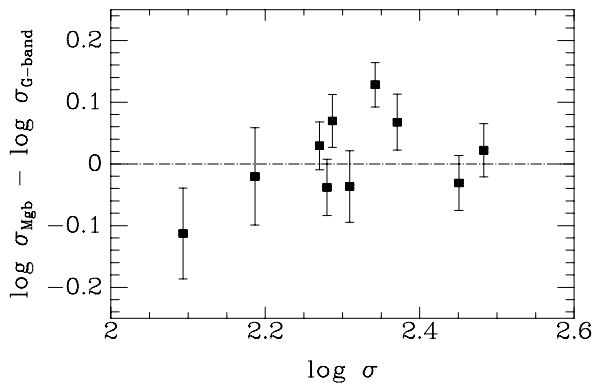}
\caption[ ]{The difference between velocity dispersions derived
from the Mgb region (rest frame 4800-5310\AA ) and from the
G-band regon (rest frame 4080-4700\AA ) versus the velocity
dispersions derived from the full wavelength range.
Only data for galaxies in A2218 are shown.
The median offset in $\log \sigma$ is $0.00\pm 0.02$.
\label{fig-compMgbG} }
\end{figure}

For the A2218 galaxies we experimented with fitting the spectra
in a limited wavelength range around the G-band region (rest frame
4080-4700\AA ) and 
around the Mgb region (rest frame 4800-5310\AA ), as well as the 
full available wavelength range (rest frame 4080-5310\AA ).
Fig.\ \ref{fig-compMgbG} shows the comparison of the velocity
dispersions derived from the G-band region and from the Mgb region.
The median offset in $\log \sigma$ is $0.00\pm 0.02$.
In the following the results from fitting the full wavelength
range are used.

The velocity dispersions were aperture corrected to a circular
aperture with diameter 1.19\,h$^{-1}$\,kpc, equivalent to
3.4 arcsec at the distance of the Coma cluster.
We adopt the technique for the aperture correction established by
J{\o}rgensen et al.\ (1995b).
The applied aperture corrections are 0.0215 and 0.028 in $\log \sigma$
for A665 and A2218, respectively.
Table \ref{tab-spec} gives the raw determinations of the 
velocity dispersions as well as the aperture corrected values.

The velocity dispersion measurements for A665 and A2218 are based
on two independent observing runs. Further, in the analysis
(Sect.\ 5) we will use the data together with measurements from
other independent observing runs. We have no way of testing
the consistency of the velocity dispersion measurements from
the different runs, since there are no galaxies in common
between the runs.
For nearby galaxies (redshifts less than 0.1) it is in general
found that velocity dispersion measurements from different runs can 
deviate with 0.01 to 0.025 in $\log \sigma$ (e.g., JFK1995b;
Smith et al.\ 1997; Wegner et al.\ 1998).
We adopt the mean offset of 0.023 in $\log \sigma$ found by
Wegner et al.\ in their analysis of 33 observing runs to be the
typical systematic uncertainty for velocity dispersion 
measurements from independent observing runs.
The uncertainty on the applied aperture correction is expected
to be significantly smaller than this systematic uncertainty.

\begin{table}
\caption{Spectroscopic parameters \label{tab-spec} }
\begin{tabular}{lrrrrr}
Galaxy    & $\log \sigma$ & $\log \sigma$ & 
            $\sigma _{\log \sigma}$ & \multicolumn{1}{c}{$z$} & S/N \\ 
          & (obs)  & ($3\farcs 4$) & & & \\ \hline
A665-1150 & 2.447 & 2.468 & 0.019 & 0.1831 & 54.5 \\
A665-1168 & 2.356 & 2.378 & 0.019 & 0.1785 & 54.5 \\
A665-1196 & 2.334 & 2.355 & 0.028 & 0.1691 & 32.6 \\
A665-2105 & 2.127 & 2.149 & 0.035 & 0.1699 & 40.7 \\
A665-2108 & ... & ...  & ...  & 0.1939 & 16.5 \\
A665-2111 & 2.369 & 2.391 & 0.025 & 0.1860 & 37.2 \\
A665-2144 & 2.267 & 2.288 & 0.034 & 0.1729 & 29.4 \\
A2218-L118 & 2.280 & 2.308 & 0.023 & 0.1748 & 32.2 \\
A2218-L148 & 2.371 & 2.399 & 0.023 & 0.1833 & 27.6 \\
A2218-L198 & 2.309 & 2.337 & 0.027 & 0.1737 & 22.4 \\
A2218-L244 & 2.287 & 2.315 & 0.020 & 0.1772 & 32.2 \\
A2218-L341 & 2.451 & 2.479 & 0.023 & 0.1630 & 25.2 \\
A2218-L391 & 2.483 & 2.511 & 0.022 & 0.1731 & 38.1 \\
A2218-L430 & 2.093 & 2.121 & 0.035 & 0.1810 & 21.7 \\
A2218-L482 & 2.342 & 2.370 & 0.018 & 0.1788 & 35.2 \\
A2218-L535 & 2.270 & 2.298 & 0.021 & 0.1809 & 31.1 \\
A2218-BOW33 & 2.186 & 2.214 & 0.038 & 0.1734 & 18.2 \\
A2218-BOW97 & ... & ... & ... & 0.291 & 12.7 \\ \hline
\end{tabular}

Notes. (obs) -- raw measurements. ($3\farcs 4$) --
aperture corrected to a circular aperture with
diameter 1.19\,h$^{-1}$\,kpc, equivalent to 3\farcs4 at the distance
of the Coma cluster.
$\sigma _{\log \sigma}$ -- formal uncertainty from the
Fourier Fitting Method. S/N -- signal-to-noise per \AA ngstr\"{o}m
in the wavelength interval used for the Fourier fitting.
Identifications for A2218 are from Le Borgne et al.\ (1992) and 
Butcher et al.\ (1983).
\end{table}

\section{Data for other clusters}

\subsection{The Coma cluster and the HydraI cluster}
 
The Coma cluster is in the following used as the low redshift reference
for our study of changes in the FP as a function of the redshift.
We use the sample of E and S0 galaxies, which has photometry in Gunn $r$
from J{\o}rgensen et al.\ (1995a), see also J{\o}rgensen \& Franx 
(1994). The sample covers the central $64'\times 70'$ of the cluster.
Velocity dispersions published by
Davies et al.\ (1987), Dressler (1987), and Lucey et al.\ (1991)
have been calibrated to a consistent
system and aperture corrected to a circular aperture with diameter 
1.19\,h$^{-1}$\,kpc, equivalent to 3.4 arcsec at the distance of the 
Coma cluster, see J{\o}rgensen et al.\ (1995b).
Further, we use new velocity dispersion measurements from
J{\o}rgensen (1999).
A total of 116 galaxies have available photometry and spectroscopy.
The sample is 93\% complete to a magnitude limit of $\mT = 15\fm 05$
in \mbox{Gunn $r$}, 
equivalent to $\MrT = -20\fm 75$ (for a distance modulus of
$35\fm 8$, $\Ho50$ and $q_{\rm o}=0.5$).

For the purpose of testing for possible differences in the coefficients
of the FP we have also used data for galaxies in the HydraI cluster.
The main purpose of including the HydraI cluster is to add more
high luminosity galaxies to the nearby sample.
We use the sample of E and S0 galaxies by 
Milvang-Jensen \& J{\o}rgensen (1999). The sample covers the
central $68'\times 83'$ of the cluster. 
A total of 45 galaxies have photometry (in Gunn $r$) and spectroscopy. 
The sample is 80\% complete to 
$\mT = 14\fm 5$ in Gunn $r$ (equivalent to $\MrT = -20\fm 0$,
for a distance modulus of $34\fm 5$).
For the galaxies brighter than $\MrT = -20\fm 75$, 
the FP for the HydraI cluster is the same as for the Coma cluster.

\begin{figure*}
\epsfxsize=16.5cm
\hspace*{0.5cm}
\epsfbox{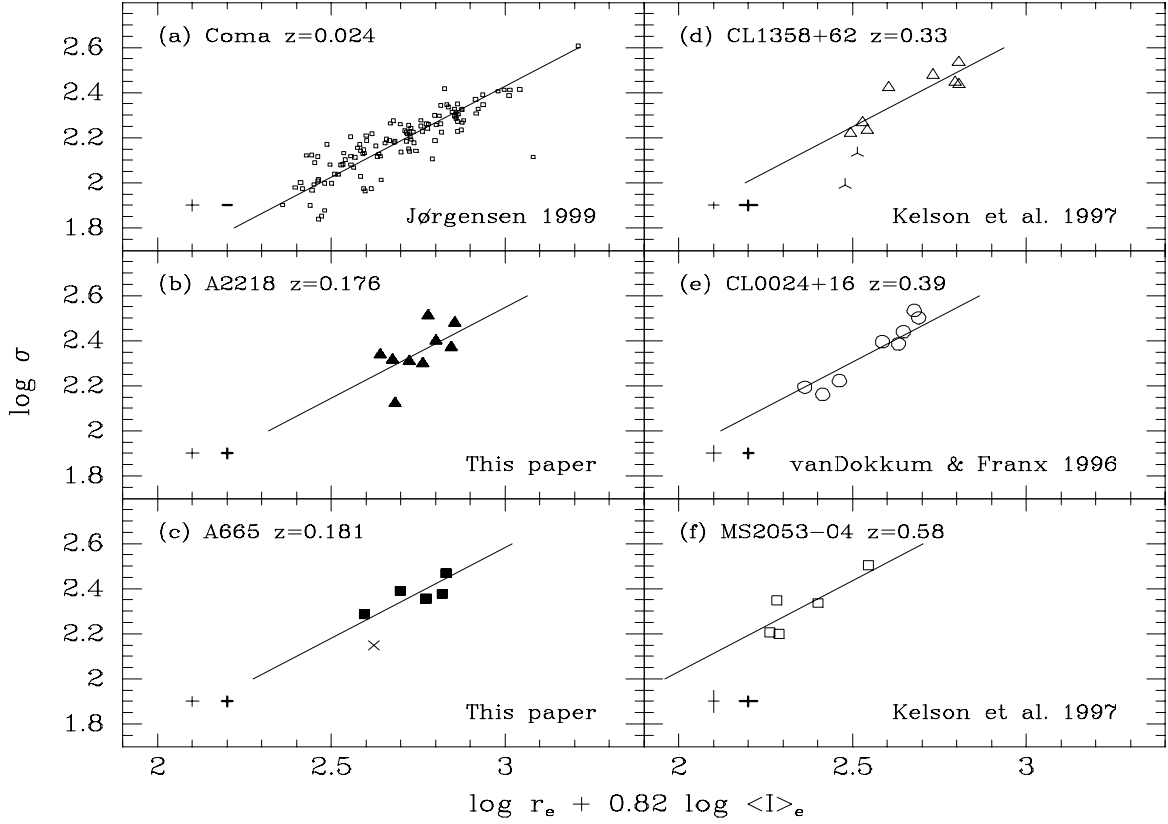}

\caption[]{The FP edge-on for Coma, A2218, A665, CL1358+62, CL0024+16,
and  MS2053-04.
The sources of the data are given on the panels.
The skeletal symbols on panel (c) and (d) are the E+A galaxies.
The photometry is calibrated to Gunn $r$ in the rest frames of
the clusters. 
The mean surface brightness $\log \Ie = -0.4(\mue -26.4)$ 
is in units of $\rm L_{\odot} / pc^2$.
The photometry is not corrected for the dimming
due to the expansion of the Universe. The effective radii are
in kpc ($\Ho50$ and $q_{\rm o} = 0.5$).
The solid lines are the FPs with coefficients adopted from JFK96
and zero points derived from the data presented in the figure.
Typical error bars are given on the panels; the thin and thick error 
bars show the random and systematic uncertainties, respectively.
\label{fig-FP} }
\end{figure*}

\subsection{Intermediate redshift clusters}

In the following analysis we also use data for the three intermediate
redshift clusters CL0024+16 (van Dokkum \& Franx 1996), 
CL1358+62 and MS2053-04 (Kelson et al.\ 1997).
Van Dokkum \& Franx and Kelson et al., who previously
studied these clusters, give effective radii, magnitudes $\Vz$
calibrated to the V-band in the rest frame of the clusters, colors,
and velocity dispersions.
We establish the transformations from the observed $\Ic$ magnitudes
and $(\Rc - \Ic )$ to Gunn $g$ in the rest frames of the clusters,
using the technique described in Section 2.3.
We then use the relation $(V-g)=-0.517(g-r)-0.027$ 
(J{\o}rgensen 1994), based on standard stars, to transform
$(\Vz - g_z )$ to $(g_z - \rz )$, where $g_z$ and $\rz$ are
Gunn $g$ and Gunn $r$, respectively, in the rest frames of the clusters.
This results in calibrations that allow us to transform from
the published magnitudes $\Vz$ and the color $(\Rc - \Ic )$ to
Gunn $r$ in the rest frames of the clusters.
For CL0024+16 at a redshift of 0.39 we find
\begin{equation}
\label{eq-cl0024}
\Vz - \rz = 0.58 (\Rc - \Ic ) -0.21
\end{equation}
For CL1358+62 at $z=0.33$ we find
\begin{equation}
\Vz - \rz = 0.50 (\Rc - \Ic ) -0.19
\end{equation}
The data for CL1358+62 include two E+A galaxies (Kelson et al.).
For MS2053-04 at $z=0.58$ we find
\begin{equation}
\label{eq-ms2053}
\Vz - \rz = 0.30 (\Rc - \Ic ) -0.15
\end{equation}

The uncertainties on $\rz$ derived using these transformations
consist of the uncertainty due to the transformation between
$(V-g)$ and $(g-r)$ (0.015) and the uncertainty due to
the transformation to $g_z$ ($0\fm 02$), a total of $0\fm 025$.
We do not include additional uncertainty from the authors'
original transformation to $\Vz$ since that uncertainty
is closely correlated with the uncertainty on the
transformation to $g_z$.
The uncertainty of $0\fm 025$ does not include the uncertainties
in the standard calibrations to $\Ic$ and $\Rc$ established by
van Dokkum \& Franx and Kelson et al.

The typical values of $\Vz - \rz$ derived from equations
(\ref{eq-cl0024}) to (\ref{eq-ms2053}) are 0.2.
This is comparable to the average $(V-r)$ color of nearby E and S0 
galaxies (e.g, J\o rgensen et al.\ 1995a).
Others found similar small color evolution
for E galaxies in clusters with redshifts less than 0.6,
e.g., Rakos \& Schombert (1995) who used passbands shifted to 
match the redshift of the studied clusters.

\begin{figure*}
\epsfxsize=18cm
\epsfbox{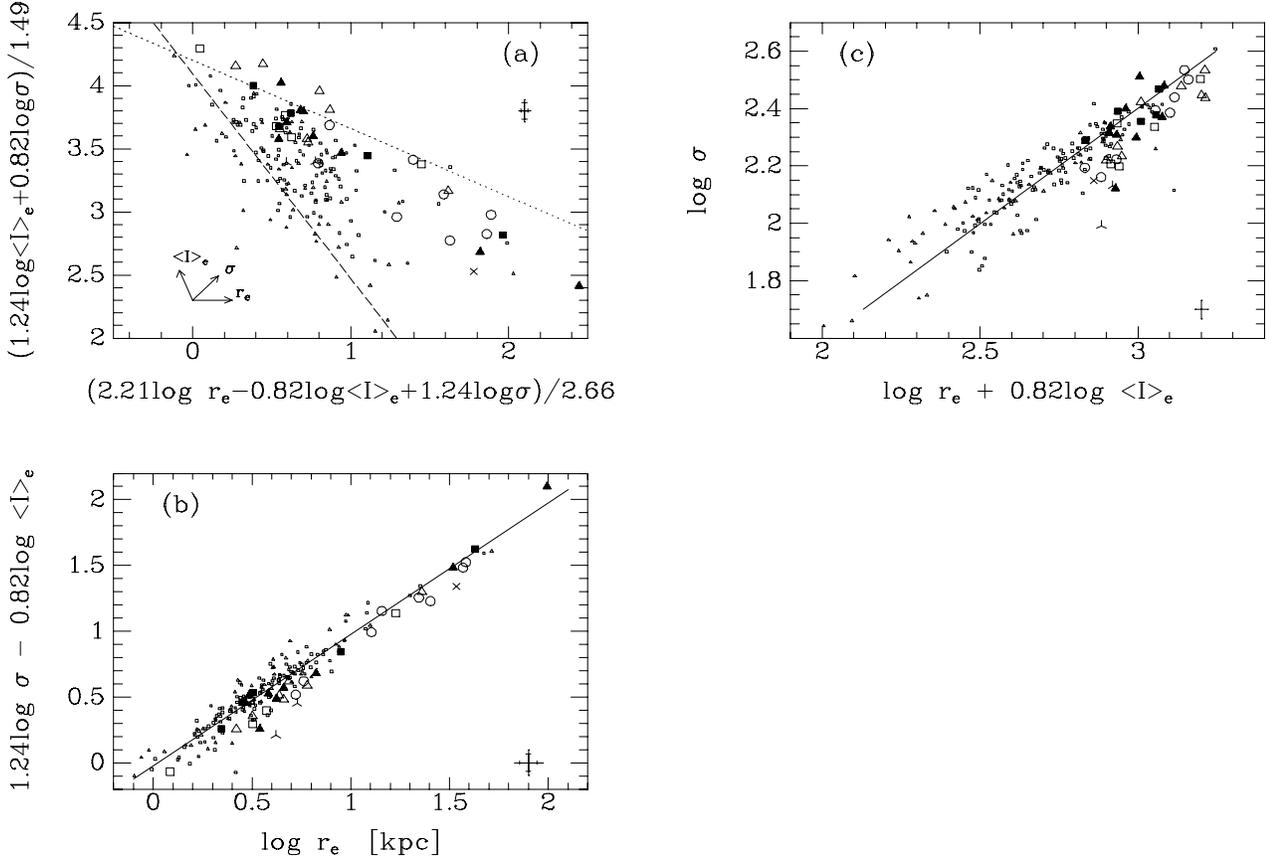}

\caption[]{The FP for Coma, HydraI and the five intermediate redshift clusters.
Small triangles -- galaxies in HydraI; other symbols as on 
Fig.\ \ref{fig-FP}.
The photometry is calibrated to Gunn $r$ in the rest frames of
the clusters and corrected for the dimming
due to the expansion of the Universe. The effective radii are
in kpc. $\Ho50$ and $q_{\rm o} =0.5$ were used.
(a) The FP face-on.
The dashed line marks the luminosity limit for the completeness of
the Coma cluster sample, $\MrT = -20\fm 75$.
The dotted line is the so-called exclusion zone for nearby galaxies, 
$y \approx -0.54x + 4.2$, first noted by Bender et al.\ (1992).
(b) and (c) The FP edge-on. The solid line is the FP for the Coma 
cluster with coefficients adopted from JFK96.
The small error bars on 
panels (a) and (b) refer to Coma and HydraI, while the larger error bars
refer to the intermediate redshift clusters.
The error bars on panel (c) refer to Coma and HydraI. The error bars
for the intermediate redshift clusters cannot be summarized in a
simple fashion, and we refer to Fig.\ \ref{fig-FP} for their sizes.
\label{fig-FP3} }
\end{figure*}

\section{The Fundamental Plane}

Fig.\ \ref{fig-FP} shows the FP edge-on for A665 and A2218
as $\log \sigma$ versus $\log \re + 0.82 \log \Ie$.
The mean surface brightness $\log \Ie = -0.4(\mue -26.4)$ 
is in units of $\rm L_{\odot} / pc^2$.
The figure also shows the FP for the Coma cluster and for 
the three intermediate redshift clusters CL1358+62 ($z=0.33$), 
CL0024+16 ($z=0.39$) and MS2053-04 ($z=0.58$).
The E+A galaxies in A665 and CL1358+62 are excluded from the following
analysis, but are shown on the figures.

We adopt the coefficients for the FP from 
JFK96, ($\alpha$,$\beta$)=($1.24\pm 0.07$,$-0.82 \pm 0.02$).
The coefficients were derived for a sample of 226 galaxies in 10 
nearby clusters.  Photometry in Gunn $r$ was used.
The relations shown on Fig.\ \ref{fig-FP} are FPs with
these coefficients.

The error bars on Fig.\ \ref{fig-FP} showing the systematic errors
include the systematic uncertainties in the photometry (cf.\ Table
\ref{tab-unphot}) and a 2\% systematic uncertainty in $\re \Ie^{0.82}$
due to any deviations from $r^{1/4}$ profiles and possible differences 
originating from differences in the fitting methods for the nearby
and intermediate redshift galaxies (cf.\ Kelson et al.\ 1997).
The systematic uncertainties on $\log \sigma$ is shown as
0.023 for the intermediate redshift clusters and zero for the 
Coma cluster, since the relevant uncertainty is the possible
inconsistency between the velocity dispersions for Coma galaxies 
and the intermediate redshift galaxies.

Fig.\ \ref{fig-FP3} shows the FP face-on and in two edge-on views. 
The HydraI sample is included on this figure.
The mean surface brightnesses have been corrected for the
dimming due to the expansion of the Universe.
The effective radii are in kpc ($\Ho50$ and $q_{\rm o} =0.5$).

The scatter around the FP is similar for A665 and A2218
(0.060 and 0.112 in $\log \re$) and for the Coma cluster (0.092 in 
$\log \re$).
Using the photometry calibrated to Gunn $r$, we find the scatter
for CL1358+62, CL0024+16 and MS2053-04 to be
0.066, 0.064 and 0.076, respectively (see also van Dokkum \& Franx 1996;
Kelson et al.\ 1997).
This is not significantly different from the scatter 
for Coma, A665 and A2218.
The scatter for the Coma cluster reported here is somewhat larger
than previous estimates based on smaller samples (e.g.,
J{\o}rgensen et al.\ 1993, 1996; Lucey et al.\ 1991).

\subsection{The coefficients of the FP}

The sample of intermediate redshift galaxies is heavily 
biased towards intrinsically bright galaxies.
The faintest galaxies observed in the intermediate reshift 
clusters typically have $\MrT = -21\fm 65$.
The selection effect is clearly visible in the face-on view of the FP, 
Fig.\ \ref{fig-FP3}a.
Further, the edge-on view on Fig.\ \ref{fig-FP3}c shows that
the samples are limited in both the velocity dispersions and in
$\log \re + 0.82 \log \Ie$.
The selection effects make it difficult to test for possible 
differences between the coefficients of the FP for Coma and the FP 
for the intermediate redshift clusters.

\begin{figure}
\epsfxsize=8.8cm
\epsfbox{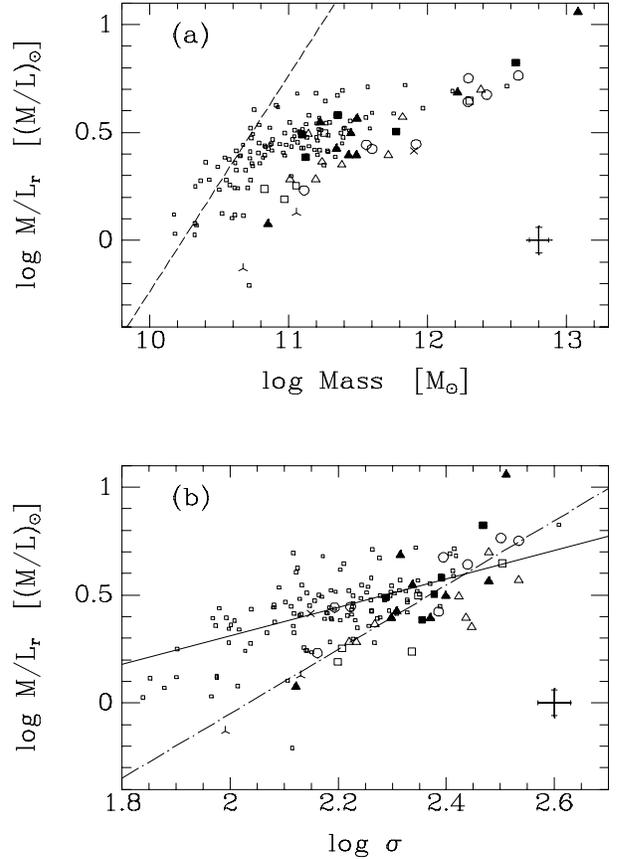}

\caption[]{(a) The M/L ratio as function of the mass.
The dashed line marks the luminosity limit for the completeness of
the Coma cluster sample, $\MrT = -20\fm 75$.
(b) The M/L ratio as function of the velocity dispersion.
Solid line -- relation fitted to the Coma cluster;
dotted-dashed line -- relation fitted to the intermediate 
redshift clusters.
Symbols as in Fig.\ \ref{fig-FP}.
The small and large error bars refer to the random uncertainties
affecting the Coma cluster and the intermediate redshift clusters, 
respectively.
\label{fig-ML} }
\end{figure}

Fig.\ \ref{fig-ML}a shows the FP as the relation between the M/L
ratio and the mass.
The masses were derived as $M = 5\re \sigma^2 /G$ 
(cf., Bender, Burstein \& Faber 1992).
This figure and Fig.\ \ref{fig-FP3}c indicate that the 
coefficient $\alpha$ for the $\log \sigma$ term in the FP is different 
for the intermediate redshift clusters and for the Coma cluster.
This was also noted by van Dokkum \& Franx (1996) in their analysis
of the CL0024+16 observations. 

We tried to fit the FP to the intermediate redshift clusters 
individually, in order to test if the coefficients of the FP for the 
intermediate redshift clusters are different from those for nearby 
clusters.  A total of 35 galaxies were included in the fits.
The E+A galaxies in A665 and CL1358+62 were omitted, since we
aim at fitting the FP for the E and S0 galaxies, only.
The fits were determined by minimization of the sum of the absolute
residuals perpendicular to the relation.
The uncertainties were derived using a bootstrap method (cf., JFK96).
All the fits gave $\beta$ values in the interval $-0.75$ to $-0.81$.
The $\alpha$ values are in the interval 0.54--1.74 and, due to
the small sample sizes, the uncertainties are large. 
We therefore repeated the fits with $\beta=-0.79$ for all the clusters.
This gives $\alpha$ in the interval 0.45--1.42, 
with uncertainties of 0.14--0.70.
%
No significant change of $\alpha$ with redshift was detected.
It must await larger samples of galaxies to map the possible change
of the coefficients as a function of the redshift.

Next, we fit the FP to the five clusters as parallel planes, 
under the assumption that the FPs for these clusters have the same 
coefficients.
For the photometry calibrated to Gunn $r$, this gives 
($\alpha$,$\beta$)=($0.89\pm 0.17$,$-0.79 \pm 0.04$).
In Johnson V we find
($\alpha$,$\beta$)=($0.92\pm 0.13$,$-0.79 \pm 0.04$).
A fit to the Coma cluster galaxies alone gives 
($\alpha$,$\beta$)=($1.27\pm 0.08$,$-0.81 \pm 0.03$),
for photometry in Gunn $r$.
The difference between $\alpha$ for the Coma cluster sample and 
$\alpha$ for the intermediate redshift sample is 0.38.
Of the 1000 bootstrap fits derived for the FP for the intermediate
redshift sample, 3.8 percent have $\alpha$ values that deviate from
$\alpha =0.89$ with more than 0.38.
Hence, the difference in $\alpha$ is significant at the 96 percent
level.

In order to explore the effect of the different selection criteria
we repeated the fit to the Coma cluster galaxies first enforcing
a magnitude limit of $\MrT = -21\fm 65$, and next enforcing limits of 
$\log \sigma \ge 2.12$ and $\log \re + 0.82 \log \Ie \ge 2.80$. 
Applying the magnitude limit, does not give a result
significantly different from the fit for the full sample,
we find ($\alpha$,$\beta$)=($1.30\pm 0.17$,$-0.82 \pm 0.06$).
The limits in $\log \sigma$ and $\log \re + 0.82 \log \Ie$ result in
($\alpha$,$\beta$)=($1.09\pm 0.27$,$-0.85 \pm 0.05$).
Though this is formally not different from the fit to the full sample,
the smaller value of the $\alpha$ coefficient indicates that
part of the difference in $\alpha$ between the Coma cluster sample
and the sample in the intermediate redshift clusters may be due to
selection effects.

\begin{figure*}
\epsfxsize=18cm
\epsfbox{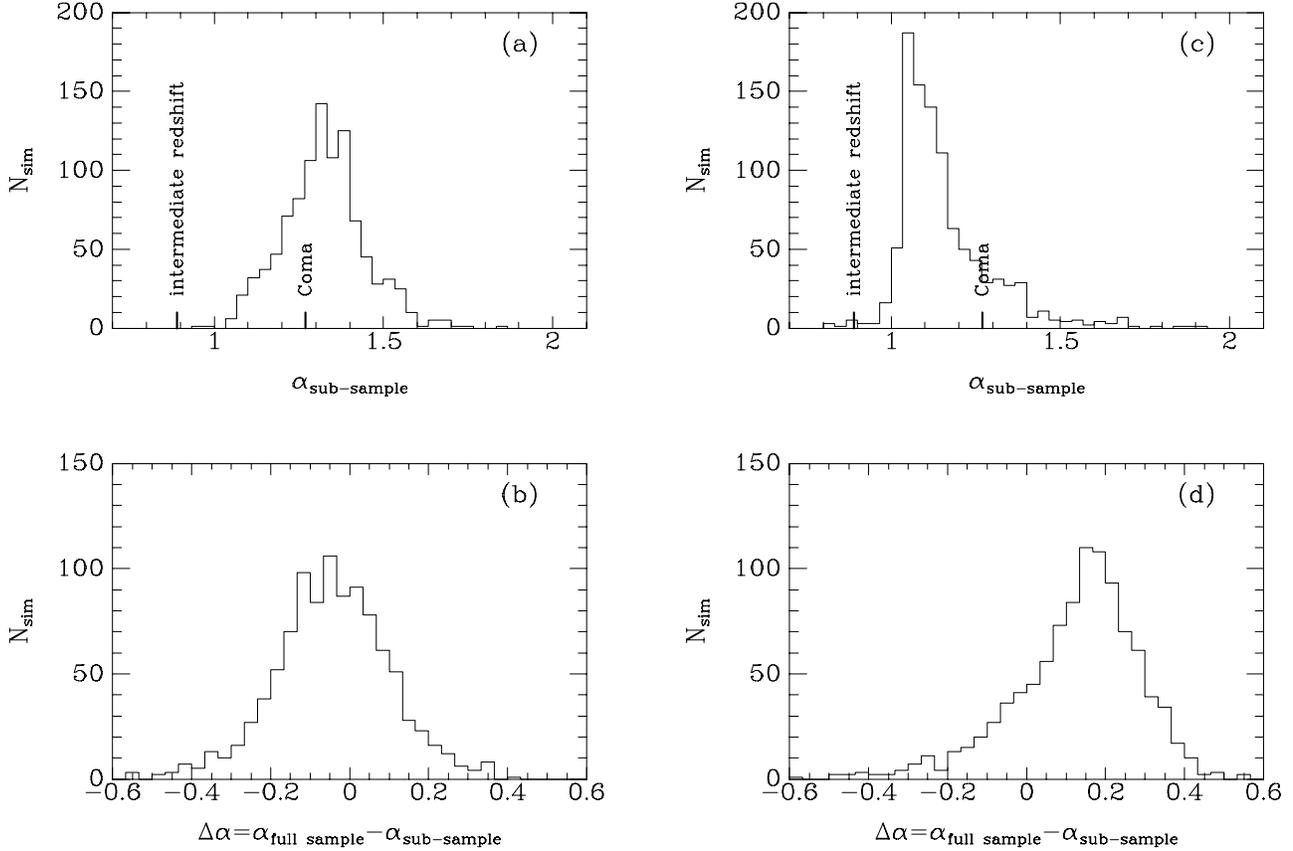}

\caption[]{(a) Histogram of the values of the coefficient $\alpha$ of 
the FP for 1000 random sub-samples of galaxies selected from
the Coma cluster and HydraI cluster sample.
The sub-samples are selected so their luminosity distributions
match the luminosity distribution for the intermediate redshift sample.
The values found for the intermediate redshift sample
and for the full Coma cluster sample are also shown.
(b) Histogram of the difference between the $\alpha$ values derived by
bootstrapping the full Coma sample and the $\alpha$ values derived
from the random sub-samples selected to match the luminosity 
distribution.
The absolute difference $|\Delta \alpha|$ is larger than 0.38 for
1.8 percent of the cases.
(c) Histogram of the values of the coefficient $\alpha$ of 
the FP for 1000 random sub-samples of galaxies selected from
the Coma cluster and HydraI cluster sample.
The sub-samples are selected so their distribution in 
$\log \re + 0.82 \log \Ie$ match the distribution for the 
intermediate redshift sample.
The values found for the intermediate redshift sample
and for the full Coma cluster sample are also shown.
(b) Histogram of the difference between the $\alpha$ values derived by
bootstrapping the full Coma sample and the $\alpha$ values derived
from the random sub-samples selected to match the 
distribution in $\log \re + 0.82 \log \Ie$.
The absolute difference $|\Delta \alpha|$ is larger than 0.38 for
3.8 percent of the cases.

\label{fig-subsamp} }
\end{figure*}

To further test if the difference in $\alpha$ may be due to selection
effects, we have performed two simulations that are both based on
fitting the FP to random sub-samples of galaxies
selected from the Coma cluster sample plus the HydraI cluster sample.
The sub-samples each contain 35 galaxies in order to match the
size of the intermediate redshift sample.
In the first simulation, the sub-samples were selected such that 
their luminosity distributions match the luminosity
distribution for intermediate redshift cluster sample.
In the second simulation, the sub-samples were selected to match
the distribution of $\log \re + 0.82 \log \Ie$.
The mean luminosity evolution of the galaxies in the intermediate
redshift clusters (see Section \ref{sec-MLevol}) was taken into account
in both simulations.
However, the results do not depend on whether this correction 
is applied.

We have included the HydraI cluster sample in these simulations 
because the Coma cluster sample has only six galaxies brighter 
than $\MrT = -23\mag$.
Our sample of galaxies in the intermediate redshift clusters
include 13 galaxies brighter than $\MrT = -23\mag$.
Thus, using only the Coma cluster sample for the test could
potentially bias the result, since the bright galaxies would
appear several times in each sub-sample.
The combined Coma and HydraI sample contains 10 galaxies brighter
than $\MrT = -23\mag$. While this may not be enough to eliminate
all bias due to some of the bright galaxies being included more
than once in each sub-sample, we expect the bias to be fairly small.

Figs.\ \ref{fig-subsamp}a and c show the distribution of $\alpha$ 
values for 1000 random sub-samples for each of the two simulations.  
The values for the full Coma cluster sample and for the intermediate 
redshift sample are overplotted.
Eight (0.8 percent) of the sub-samples selected to match the 
luminosity distribution result in values of 
$\alpha$ within the one sigma error bar of the coefficient found 
for the intermediate redshift sample.
For the sub-samples selected to match the distribution in 
$\log \re + 0.82 \log \Ie$, 20.3 percent give $\alpha$ values within the
one sigma error bar.

The distributions of the $\beta$ values for the sub-samples 
encompass for both simulations the value for the Coma cluster 
sample as well as the value for the intermediate redshift sample, 
which in turn are not significantly different from each other.

In order to construct two-sided tests, we compare the $\alpha$ 
values derived from the two simulations of the sub-samples 
to those derived by bootstrapping the full Coma sample. 
The comparison with the first simulation gives a measure of how 
often we can expect 
two samples with luminosity distributions similar to the intermediate 
redshift sample and the Coma cluster sample, respectively, 
and drawn from the same underlying distribution, to give a 
difference in $\alpha$ larger than the measured difference of 0.38.
The comparison with the second simulation gives the same measure 
for samples selected on the distribution in $\log \re + 0.82 \log \Ie$.
Fig.\ \ref{fig-subsamp}b and d show the distributions of the $\alpha$
difference for the two comparisons. 
For the sub-samples selected to match the luminosity distribution,
18 of the 1000 cases, or 1.8 percent, 
have an absolute difference $|\Delta \alpha|$ larger than 0.38.
For the sub-samples selected to match the distribution in 
$\log \re + 0.82 \log \Ie$, 3.8 percent of the cases have
$|\Delta \alpha| > 0.38$. 

Based on these simulations,
we conclude that the difference in the $\alpha$ coefficient
of the FP is not likely to be due to the difference in the luminosity
distributions of the two samples.
It is possible that part of the difference in the $\alpha$ coefficient
is due to the distribution of $\log \re + 0.82 \log \Ie$ being
different for the two samples.
The different appearances of the distributions shown on 
Figs.\ \ref{fig-subsamp}a and c add to the evidence that
the selection effect present in $\log \re + 0.82 \log \Ie$ may
affect the derived value of $\alpha$.

\begin{table*}
\begin{minipage}{12.9cm}
\caption[]{Uncertainties on the FP zero point \label{tab-uncFP} }
\begin{tabular}{lrrrl}
       & \multicolumn{3}{c}{\rule[1mm]{9mm}{0.2mm} Clusters \rule[1mm]{9mm}{0.2mm}} \\
Source & Coma & A665 & A2218 & random/systematic \\ \hline
Photometric zero point $\sigma(\mue)$ & 0.015 & 0.02  & 0.02  & systematic \\
Velocity dispersion                   & 0.000 & 0.023 & 0.023 & systematic \\
Coma peculiar velocity, 300km/s, $\sigma(\log \re)$ & 0.018 & ... & ... & systematic \\
Total systematic uncertainty, $\sigma(\log \re)$ & 0.023 & 0.035 & 0.035 & systematic \\
Uncertainty on the FP zero point$^a$, $\sigma (\log \re)$ & 0.009 & 0.027 & 0.037 & random \\ \hline
\end{tabular}

Note. The uncertainty on the FP zero point is $\sigma _{\rm fit}/N^{1/2}$ 
where $\sigma _{\rm fit}$ is the rms scatter of the FP and $N$ is the number of 
observed galaxies.
\end{minipage}
\end{table*}

The coefficients we find for the FP for the intermediate redshift 
clusters imply that
\begin{equation}
M/L \propto \re ^{0.27} \sigma ^{0.87} \propto M^{0.44} \re^{-0.17}
\end{equation}
This should be compared with 
$M/L \propto \re ^{0.22} \sigma ^{0.49} \propto M^{0.24} \re^{-0.02}$
found for nearby clusters (JFK96).
The difference may indicate that the low mass galaxies show a stronger 
luminosity evolution than the more massive galaxies.
The driving factor may however be the velocity dispersion rather
than the mass of the galaxies.
A direct fit to the M/L ratio as a function of the velocity dispersion
gave for the nearby clusters $M/L \propto \sigma ^{0.86}$ (JFK96).
Fig.\ \ref{fig-ML}b shows the M/L ratio versus the velocity
dispersion for the Coma cluster and the intermediate redshift clusters.
The different slopes for the two samples is clearly seen.
For the Coma cluster alone we find $M/L \propto \sigma ^{0.66\pm 0.13}$,
while the sample of galaxies in the intermediate redshift clusters
gives $M/L \propto \sigma ^{1.49 \pm 0.29}$.
The relations for the Coma cluster and for the intermediate redshift 
clusters are overplotted on Fig.\ \ref{fig-ML}b.

If real, the difference in the slopes of the FP (and the slopes of the
M/L-$\sigma$ relation) may be due to bursts of star formation in
the low mass galaxies in the intermediate redshift clusters leading
to a lower M/L ratio for these galaxies.
Alternatively, the samples in the intermediate redshift clusters
may contain some early-type spirals.

\subsection{The evolution of the M/L ratio \label{sec-MLevol} }

The zero point of the FP depends on the cosmological effects
(surface brightness dimming, the value of $q_{\rm o}$, and the
value of the cosmological constant), and the evolution of the galaxies.
We assume a cosmological constant of zero.
We adopt the coefficients for the FP for nearby 
clusters (JFK96) for all the clusters. 
Because of the possible difference in the coefficients of the FP
(cf.\ Section 5.1) the following discussion of the evolution of
the M/L ratios should be understood as the average evolution of
the available samples of galaxies.
Specifically, the results refer to galaxies with masses above 
$6.5 \cdot 10^{10} {\rm M_{\odot}}$ ($\Ho50$ and $q_{\rm o} =0.5$).

Table \ref{tab-uncFP} summarizes the sources of uncertainty 
affecting the zero points of the FP for Coma, A665 and A2218.
The total systematic uncertainty on the FP zero point is
the sum of the systematic uncertainties and is given as 
$\sigma (\log \re)$.
The random uncertainties introduced by the random photometric
uncertainties and fitting uncertainties are included in the
uncertainty on the FP zero point since this is based on the
measured rms scatter of the FP.

On the following figures the error bars reflect the uncertainties
listed in Table \ref{tab-uncFP}.
The uncertainties for the other intermediate
redshift clusters were adopted from van Dokkum \& Franx (1996)
and Kelson et al.\ (1997), and where necessary the total
uncertainties were derived in the same way as for A665 and A2218.

\begin{figure}
\epsfxsize=8.8cm
\epsfbox{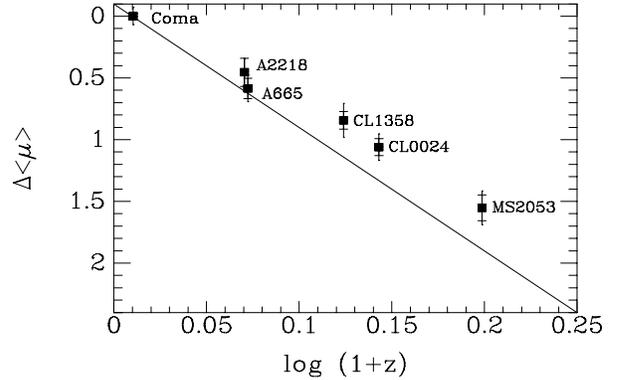}

\caption[]{The offsets in the surface brightnesses derived from the
FP versus the redshift of the clusters. $q_{\rm o}=0.5$ was used.
The small and large error bars show the random and systematic 
uncertainties, respectively. These were derived as 
$\sigma(\mue) = \sigma(\log \re) \cdot 2.5 / 0.82$. 
See Table \ref{tab-uncFP} for values of $\sigma (\log \re)$.
The solid line is the expected relation for an expanding Universe 
and no evolution of the galaxies.
The data are in agreement with an expanding Universe and a slow
luminosity evolution.
A smaller $q_{\rm o}$ results in smaller values of $\Delta \mmu$, 
and therefore a larger luminosity evolution.
\label{fig-SBT} }
\end{figure}

Fig.\ \ref{fig-SBT} shows the offsets in the surface brightnesses,
$\Delta \mmu$, based on the zero points of the FPs.
We  used $q_{\rm o}=0.5$.
No correction for the expansion of the Universe has been applied and
the solid line on the figure shows the expected relation for
an expanding Universe.
A smaller $q_{\rm o}$ results in smaller values of $\Delta \mmu$.
The  data are consistent with an expanding Universe and a slow
luminosity evolution of the galaxies.
In the following we correct the data for the expansion of the
Universe, and then use the FP to study the mean
evolution of the galaxies in the clusters.

\begin{figure}
\epsfxsize=8cm
\epsfbox{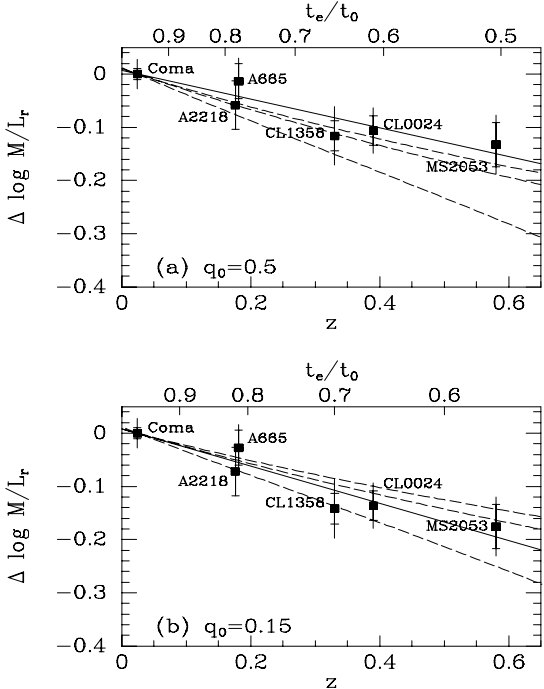}

\caption[]{The evolution of the M/L ratio as function of the redshift of
the clusters.  The photometry has been calibrated to Gunn $r$ in 
the rest frame of the clusters.
The small and large error bars show the random and systematic 
uncertainties, respectively. These were derived as 
$\sigma (\log M/L) = \sigma (\log \re) / 0.82$. 
See Table \ref{tab-uncFP} for values of $\sigma (\log \re)$.
(a) $q_{\rm o}=0.5$; (b) $q_{\rm o}=0.15$.
The top axis on each panel shows the time of emission in units 
of the present age of the Universe.
Solid lines -- least squares fits to the data.
Dashed lines -- passive evolution with $z_{\rm form}=\infty, 5$ and 1
(top to bottom).
\label{fig-MLevol} }
\end{figure}

The FP zero point differences between the intermediate redshift
clusters and the Coma cluster can be interpreted as differences
in the M/L ratios. 
With the FP coefficients from JFK96 the M/L ratios can be expressed as 
\begin{equation}
\label{eq-ML}
\begin{array}{ll}
\log M/L & = 0.49 \log \sigma + 0.22 \log \re - \gamma/0.82 + cst \\
         & = 0.24 ( 2 \log \sigma + \log \re ) - 0.02 \log \re \\
         & \hspace*{0.5cm}- \gamma/0.82 + cst 
\end{array}
\end{equation}
(cf., JFK96), where $\gamma$ is the FP zero point.
Under the assumption that no merging or accretion takes place,
the mass and therefore $( 2 \log \sigma + \log \re )$ is constant.
We can then derive the differences in the M/L ratios from the 
differences in the FP zero points, ignoring the small term
$0.02 \log \re$.
Fig.\ \ref{fig-MLevol} shows the differences in the M/L ratios as
a function of the redshifts.

The galaxies in A665 and A2218 have on average
M/L ratios ($11\pm 7$)\% smaller than the M/L ratios of the galaxies 
in the Coma cluster.
Together with the previous data from van Dokkum \& Franx (1996) and 
Kelson et al.\ (1997), this shows the gradual and slow evolution of the 
M/L ratio of the galaxies as a function of the redshift.
A least squares fit to the all data shown on Fig.\ \ref{fig-MLevol}
gives $\Delta \log M/L_r = (-0.26\pm 0.06) \Delta z$ for 
$q _{\rm o}=0.5$ and 
$\Delta \log M/L_r = (-0.34\pm 0.06) \Delta z$ for $q _{\rm o}=0.15$.
For the photometry calibrated to Johnson V we find
$\Delta \log M/L_V = (-0.29\pm 0.08) \Delta z$ and 
$\Delta \log M/L_V = (-0.37\pm 0.08) \Delta z$ for
$q _{\rm o}=0.5$ and 0.15, respectively.
Our results in Johnson V are in agreement with the results by 
van Dokkum \& Franx and Kelson et al.

The M/L ratio of a stellar population can be modeled as
\begin{equation}
M/L \propto t^{\kappa}
\label{eq-MLevol1}
\end{equation}
where $t$ is the age of the stellar population (Tinsley 1980).
The coefficient $\kappa$ depends on the initial mass function (IMF) 
of the stars and on the passband.
Based on the single burst stellar population models from 
Vazdekis et al.\ (1996), we find for photometry in \mbox{Gunn $r$} 
that $\kappa _r \approx 0.9-0.2x$ where $x$ is the slope of the IMF.
A Salpeter (1955) IMF has $x=1.35$ (equivalent to $\kappa _r \approx 0.6$).
For photometry in Johnson V, the expected evolution is nearly the same.
The models from Vazdekis et al.\ give $\kappa _V \approx 0.95-0.2x$.

Equation (\ref{eq-MLevol1}) can be used to relate the change in
the  M/L ratio with redshift
to the formation redshift, $z_{\rm form}$, and the value of $q_{\rm o}$.
The dashed lines on Fig.\ \ref{fig-MLevol} show the expected relations
for $\kappa =0.6$ and $z_{\rm form}=\infty, 5$ and 1.
Models with shallower IMFs make the slope of the expected relation
steeper, while a model with steeper IMFs make the slope of
the relation shallower.

For $q_{\rm o}=0.5$, the models with a Salpeter IMF and 
$z_{\rm form} > 5$ are consistent with the data within 2$\sigma$.
A formation redshift of $z_{\rm form}= 2.4$ represents a 3$\sigma$
deviation. Models with $z_{\rm form}$ smaller than 2.4 deviate more
than 3$\sigma$ from the fit to the data.
If we assume $q_{\rm o}=0.15$ more models are consistent with the data. 
A model with a Salpeter IMF then implies $z_{\rm form} \approx 2.5$.
Models with $z_{\rm form}$ in the interval 1.7--6.5 are consistent
with the data within 2$\sigma$.
Models with $z_{\rm form} < 1.4$ deviate more than 3$\sigma$ from
the fit to the data.



Because of the larger number of clusters analyzed here we are
able to put stronger formal limits on the formation redshift,
$z_{\rm form}$, than done in previous studies (van Dokkum \& Franx 1996;
Kelson et al.\ 1997).
However, the derived model constraints should only be taken as rough 
guidelines. The correct interpretation of the data is most likely 
rather more complicated than indicated here. 
The correct value of $\kappa$ is not known. Further, the evolution
of E and S0 galaxies cannot be viewed as a single burst event.
The presence of younger stellar populations in the galaxies
would imply stricter lower limits on $z_{\rm form}$.
The most fundamental assumption for the interpretation of the data
is that the observed E and S0 galaxies in the intermediate redshift
clusters in fact evolve into galaxies similar to the present day
E and S0 galaxies.
It is possible that our selection of E and S0 galaxies in the
intermediate redshift clusters is biased to select already
aged galaxies (see also Franx \& van Dokkum 1996).
In this case we are likely to underestimate the mean evolution
measured from the FP
when comparing to complete samples of galaxies in nearby clusters.
Larger and more complete samples in intermediate redshift clusters 
are needed in order to address this problem in detail.

\section{Conclusions}

We have established the Fundamental Plane (FP) for the two rich
clusters Abell 665 and Abell 2218, both at a redshift of 0.18.
The photometric parameters were derived from ground-based
observations obtained with the Nordic Optical Telescope, La Palma,
and from HST observations.
The photometry was calibrated to Gunn $r$ in the rest frame of the
clusters.
Central velocity dispersions were measured for six galaxies in
A665 and ten galaxies in A2218.
The FPs for the two clusters were compared to the FP for nearby
clusters derived by JFK96, and to the FP for the Coma cluster.
The scatter around the FP for A665 and A2218 is similar to the
scatter found for nearby clusters.

We have used the data for A665 and A2218 together with other
recently available data for intermediate redshift clusters
(van Dokkum \& Franx 1996; Kelson et al.\ 1997) to test if
the slope of the FP changes with redshift.
The data for the clusters CL0024+16 (z=0.39), CL1358+62 (z=0.33) 
and MS2053-04 (z=0.58) were 
calibrated to Gunn $r$ in the rest frame of the clusters.
We find that the coefficient for the $\log \sigma$ term in the FP is
significantly smaller for the intermediate redshift clusters than
for the Coma cluster. 
The difference cannot be explained by the difference in
the luminosity distribution of the Coma cluster sample and the
samples in the intermediate redshift clusters.
However, other types of selection effects, like a difference
in the distribution of the combination $\log \re +0.82 \log \Ie$,
have not been ruled out.
The smaller value of the coefficient implies that 
$M/L \propto M^{0.44} \re ^{-0.17}$ for the intermediate redshift 
clusters, compared to $M/L \propto M^{0.24} \re ^{-0.02}$ 
for nearby clusters (JFK96).
This may indicate that the low mass galaxies in the clusters
evolve faster than the more massive galaxies.
We cannot distinguish if this is purely a difference in the
luminosity evolution or if also dynamical evolution takes place.
Alternatively, the velocity dispersion and not the mass could be 
the driving parameter.
Direct fits to the M/L ratio as a function of the velocity
dispersion give $M/L \propto \sigma^{0.66}$ for the Coma cluster
and $M/L \propto \sigma^{1.49}$ for the intermediate redshift clusters.

The zero point offsets in the FP for the intermediate redshift clusters
were used to investigate the average evolution of the M/L ratios of
the galaxies.
The M/L ratios of the galaxies in A665 and A2218 are on average 
($11\pm 7$)\% smaller than the M/L ratios of the galaxies in the Coma 
cluster.  From all the clusters we find that the M/L ratio changes with
the redshift as $\Delta \log M/L_r = (-0.26\pm 0.06) \Delta z$
(for $q_{\rm o}=0.5$),
equivalent to a change in the absolute magnitudes of
$\Delta \MrT = (-0.65 \pm 0.15) \Delta z$.
The slow evolution of the M/L ratios as a function of redshifts
that we find is in general agreement with previous results by 
van Dokkum \& Franx (1996), Kelson et al.\ (1997), Bender et al.\ (1998)
and Pahre et al.\ (1999).
The result can be used to constrain the formation redshift of the 
galaxies.
The interpretation depends on the crucial assumption that the E and
S0 galaxies observed in the intermediate redshift clusters evolve
into galaxies similar to present day E and S0 galaxies
(cf., Franx \& van Dokkum 1996), and 
that the samples in the intermediate redshift clusters form a
representative sample of all the galaxies that will end as
present day E and S0 galaxies.
For $q_{\rm o}=0.5$, models with a single burst population,
a Salpeter IMF and $z_{\rm form} > 5$ are consistent within
2$\sigma$ with the observed evolution of the M/L ratios. 
Models with $z_{\rm form} < 2.4$ deviate more than 3$\sigma$ from
a least squares fit to the data.
For $q_{\rm o}=0.15$, we find $z_{\rm form} \approx 2.5$, while
models with $z_{\rm form} < 1.4$ deviate more than 3$\sigma$
from the fit to the data.
The model constraints presented here are stronger than in previous
studies due to the larger number of clusters included in the analysis.

\vspace{0.5cm}
Acknowledgements:
The staff at NOT, KPNO and MMT are thanked for assistance 
during the observations.
L.\ Davis is thanked for providing standard magnitudes for stars
in M92.
K.\ Kuijken and D.\ Kelson are thanked for making the template star 
observations available.
Financial support from the Danish Board for Astronomical Research
is acknowledged.
Support for this research was provided by NASA through grant
number HF-1016.01.91A to MF and grant number HF-01073.01.94A to
IJ, both grants from the Space Telescope Science Institute,
which is operated by the Association of Universities for Research
in Astronomy, Inc., under NASA contract NAS5-26555.
JH was supported in part by the Danish Natural Science Research Council.

\appendix

\section{The photometry}

\subsection{Basic reductions of ground-based images}

The bias level and pattern turned out to depend on the CCD temperature,
which in turn varied with $\pm 1.0$K.
Therefore, the bias was subtracted row-by-row based on the overscan
region of the images.
After this processing a few of the images showed row-to-row 
variations in the background of 1-1.5\%.
The level of these variations were determined directly from the 
images, and subtracted out with appropriate filtering to ensure
the levels were not affected by signal from the objects.
No correction for the dark current was performed. The dark current
is less than 2$e^-$ per hour.

The offset between the set exposure time and the actual exposure
time for the iris shutter of the camera was
determined from dome flats. The average shutter correction
is $\Delta = 0.0826\pm 0.0008$ sec.

The linearity of the CCD was tested from dome flats.
No signs of non-linearity can be detected within the detection
accuracy of 1.5\%, for levels up to the software saturation limit.

Flat fields were constructed from twilight flat fields.
Further, the low frequency variations over the fields were corrected
for by using night time exposures of blank fields.
Dust grains gave rise to flat field variations from night to night,
and sometimes during the night.
Night time exposures of blank fields were also used to correct for
this effect.
The pixel-to-pixel accuracy of the flat fields is 0.2-0.4\% in
areas not affected by the dust grains, and 0.8\% in areas
affected by the dust grains.
The low frequency variations over the field and any residuals 
from the dust grains are below 0.5\% after the flat field correction.

The I-band images of the two galaxy clusters were combined and
cleaned for signal from cosmic-ray-events (CR) as follows.
Geometrical transformations (shifts in $x$ and $y$ and a rotation)
were used to transform all images of a field to a common reference image.
The median sky level was subtracted from each image, and the
relative intensity was determined from aperture photometry of
objects brighter than $\approx 20\fm 5$ in the I-band.
A median image was constructed, each image scaled according to the
relative intensity.
Each image was then compared to the median image and deviations
larger than seven times the local noise were flagged as CRs.
Mask images of the flagged pixels as well as other bad pixels due
to the CCD were constructed.
Finally, a mean image was derived, omitting the flagged pixels,
and each individual image scaled according to the relative intensity
of the image.
To maintain the statistically properties of the mean image, the mean
of the median sky levels was added to the final image.
The masks were inspected visually to make sure no signal from objects
got masked.
Further, the cleaned mean image was compared with a mean image with
no masking by means of aperture photometry of the objects in the field.

The magnitude zero points for the combined images were established
by comparing aperture photometry from these images to aperture
photometry derived from the individual images.
Photometry was derived for 40-60 objects with
I-band magnitudes brighter than $\approx 20\fm 5$.
The consistency of the zero points is better than $0\fm 015$.

\subsection{Basic reduction of HST images}

The HST/WFPC2 images were corrected for bias, dark-current and
flat field corrected using the pipeline reduction
supplied by Space Telescope Science Institute.
The images were co-added and cleaned for cosmic-ray-events using
the iterative method implemented in the Space Telescope Science
Data Analysis Software (STSDAS) task ``crrej''.
Where necessary, the individual images were first registered
by applying small integer pixel shifts.

\subsection{Calibration to standard passbands}

\subsubsection{Ground-based photometry \label{sec-NOTstd} }

The NOT photometry was standard calibrated based on observations
of stars in M67, M92, and the field SA101 taken during the nights of the
galaxy observations. All nights during which galaxy observations 
were obtained were photometric.
Aperture photometry was derived for 188 observations of a total of 
35 stars. 
The observations have large variations in the seeing, and some
of the exposures were taken with the telescope slightly defocused
due to the brightness of the stars.
The aperture radii for the photometry were typically $4\farcs 9$.
Based on growth curves of bright stars in the fields
the aperture magnitudes were corrected for the finite sizes of the 
apertures. 
The growth curves typically extended to $9\farcs 5$, and
the typical size of the correction is $-0\fm 04$.
This correction ensures consistency between the defocused and 
focused observations, and between observations with large
differences in the seeing.
All observations in the I-band were obtained during night 1 and
night 5. All observations in the V-band were obtained during 
night 2. The typical seeing for the standard star observations
was FWHM=$0\farcs 6 - 0\farcs 7$, $1\farcs 5 - 2\farcs 5$ and
$0\farcs 7 - 1\farcs 0$ for the nights 1, 2 and 5, respectively.

The 9-15 stars in M67 were used for determination of the 
atmospheric extinction.
We find $k_I ({\rm night 1}) = 0.063 \pm 0.001$,
$k_I ({\rm night 5}) = 0.124 \pm 0.001$, and
$k_V ({\rm night 2}) = 0.165 \pm 0.004$.
Further, observations from night 5 were offset to consistency with
night 1 by subtraction of $0\fm011$.

We have used standard magnitudes from Landolt (1992) (SA101) and
Davis (1994) (M92).
The magnitudes from Davis are consistent with Christian et al.\
(1985) and Stetson \& Harris (1988) for the stars in common.
The standard magnitudes from M67 are from Montgomery et al.\ (1993),
Joner \& Taylor (1990) and J{\o}rgensen (1994), in this 
priority. The magnitudes from the three sources are consistent for
the stars in common.
We derived the following standard transformations
\begin{equation}
\label{eq-std}
\begin{array}{lll}
\Ic = I_{\rm inst} + {\rm cst} & {\rm rms=0.021} \\
V = V_{\rm inst} + 0.0596 (V-I)_{\rm inst} + {\rm cst} & {\rm rms=0.014}
\end{array}
\end{equation}
The I magnitudes were calibrated to Cousins $\Ic$.
The transformation for $\Ic$ has no significant color term;
if we include a $(V-I)$ term in the transformation we find
a coefficient for this term of $0.001\pm 0.007$.
The photometry for the galaxies was corrected for the galactic
extinction.
For A665, the galactic extinction is $A_{\rm B} = 0.15$ 
equivalent to $A_{\rm I} =0.07$ and $A_{\rm V} =0.11$.
A2218 has a galactic extinction in the B-band of $A_{\rm B} = 0.10$,
equivalent to $A_{\rm I} =0.05$ and $A_{\rm V} =0.08$.
The galactic extinctions were derived from the reddening maps 
published by Schlegel, Finkbeiner \& Davis (1998).

\subsubsection{Photometry from HST observations \label{sec-HSTstd} }

The photometry from the HST images was calibrated using
the transformations given by Holtzman et al.\ (1995b).
For F814W we use the transformation based on observations,
while for F606W and F702W we use the synthetic transformations.
We define instrumental magnitudes as
\begin{equation}
m_{\rm inst} = -2.5 \log ({\rm DN}/t_{\rm exp}) + {\rm ZP} + 2.5 \log {\rm GR}_i
 + \Delta m + 0.05
\end{equation}
``DN'' is the number of counts, ``ZP'' are the zero points given
by Holtzman et al.\ (1995b) for the relevant transformation
(see equations [\ref{eq-HSTVI}] and [\ref{eq-HSTR}]).
We use ${\rm GR}_2=2.003$, ${\rm GR}_3=2.006$ and ${\rm GR}_4=1.955$.
The term 0.05 is added as a correction for the difference
between ``short'' and ``long'' exposures (Hill et al.\ 1998).
The aperture corrections $\Delta m$ were estimated from data for the 
encircled energy given by Holtzman et al.\ (1995a).
We use $\Delta m_{814}=0.0983$, $\Delta m_{702}=0.1087$, and
$\Delta m_{606}=0.1138$.
The instrumental magnitudes in the three filters are then given by
\begin{equation}
\begin{array}{lcl}
m_{814} & = &  -2.5 \log ({\rm DN}/t_{\rm exp}) + 20.987 + 2.5 \log {\rm GR}_i\\
m_{702} & = &  -2.5 \log ({\rm DN}/t_{\rm exp}) + 21.670 + 2.5 \log {\rm GR}_i\\
m_{606} & = &  -2.5 \log ({\rm DN}/t_{\rm exp}) + 22.257+ 2.5 \log {\rm GR}_i\\
\end{array}
\end{equation}
The photometry for galaxies in A665 was calibrated using the
transformations for $\Ic$ and $V$ and colors based on HST data.
We use the following transformations
\begin{equation}
\begin{array}{lcl}
\Ic & = & m_{814} - 0.062 (V-\Ic) + 0.025 (V-\Ic)^2 \\
V & = & m_{606} + 0.254 (V-\Ic) + 0.012 (V-\Ic)^2 
\end{array}
\label{eq-HSTVI}
\end{equation}
The $\Ic$ transformation is for the in-flight system based 
on stars bluer than the typical colors of the observed galaxies,
$(V-\Ic )= 1.5$. However, for $(V-\Ic )= 1.5$ the transformation 
agrees within $0\fm 005$ with the synthetic transformation.

A2218 was observed in the F702W filter, only.
We have calibrated this photometry to $\Ic$ using the transformation
for $\Rc$ given by Holtzman et al.\ (1995b) and the relation
$(V-\Rc ) = 0.52(V-\Ic)$, which we derived from standard star
magnitudes for stars with $(V-\Ic)$ in the interval 1--2 (Landolt 1992).
The coefficient is in agreement with colors of E and S0 galaxies
at $z=0.2$ (e.g., Fukugita, Shimasaku \& Ichikawa 1995; 
Frei \& Gunn 1994). This gives the transformation
\begin{equation}
\Ic =  m_{702} - 0.227 (V-\Ic) - 0.021 (V-\Ic)^2
\label{eq-HSTR}
\end{equation}
We used colors from the ground-based photometry in order to calibrate
the HST observations of A2218.
Finally, the photometry was corrected for the galactic extinction.

\subsection{Photometric parameters}

\refstepcounter{table}
\label{tab-photA665}
\refstepcounter{table}
\label{tab-photA2218}

\begin{table*}
\hspace*{0.0cm}
\vbox to24cm{\rule{0pt}{18cm}}
\includegraphics{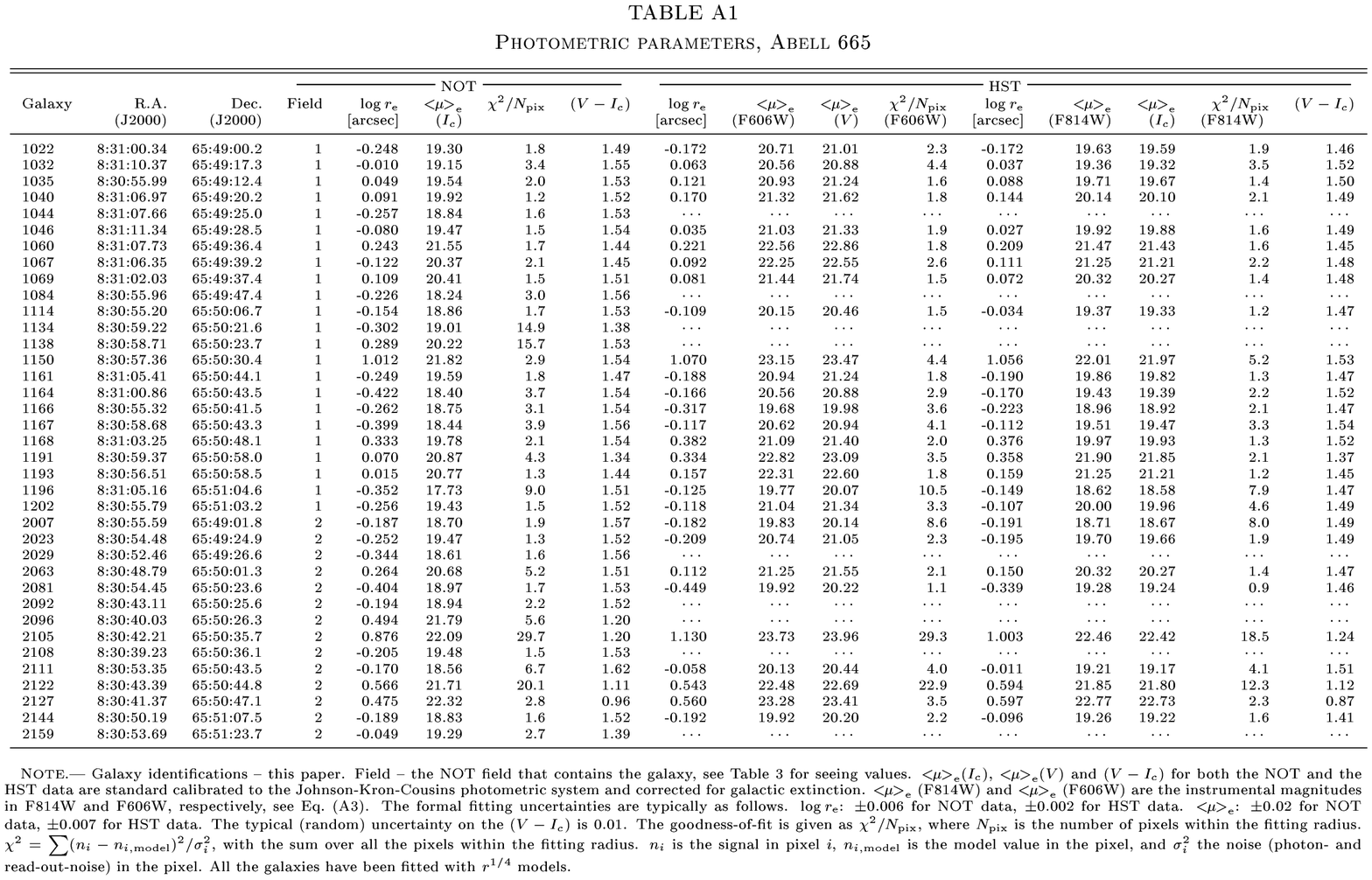}
\end{table*}

\begin{table*}
\epsfxsize=18cm
\epsfbox{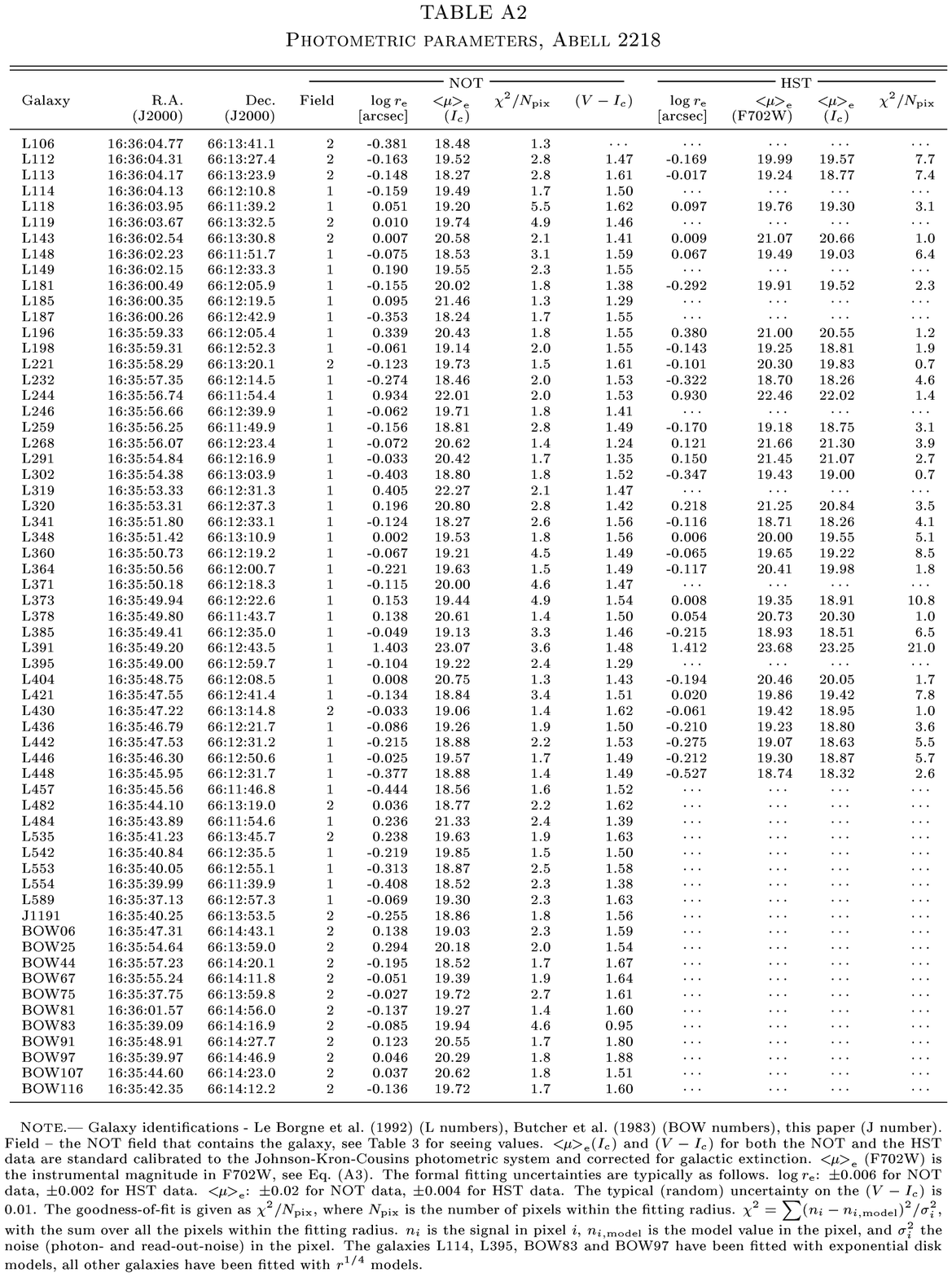}
\end{table*}

Tables \ref{tab-photA665} and \ref{tab-photA2218} give the effective
parameters and colors of all galaxies in the two clusters brighter
than $\mT = 19\mag$ in $\Ic$ and within the fields covered by
the NOT observations. 
Five of the galaxies in A665 were also observed by Oegerle et al.\ (1991).
The cross-references are as follows, with our galaxy number given first.
1032 = 218, 1150 = 201, 1168 = 224, 2105 = 225 and 2122 = 227.

\subsection{Comparison of ground-based and HST photometry}

\begin{figure}
\epsfxsize=8.8cm
\epsfbox{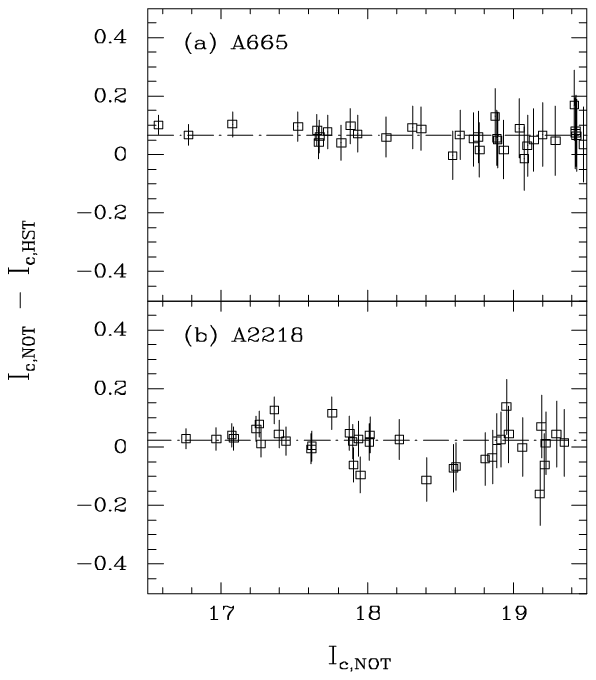}

\caption[ ]{Comparison of aperture magnitudes. 
An aperture with radius 3 arcsec was used.
\label{fig-apmag} }
\end{figure}

The ground-based photometry and the HST photometry were compared
by means of aperture magnitudes.
In order to limit the effect of the different spatial resolution
and the choice of the aperture size we convolved the HST images
to approximately the same resolution as the ground-based images.
Aperture magnitudes for the galaxies were then derived within 
apertures of with radii of 3 arcsec.
The magnitudes were standard calibrated as described in Sections
\ref{sec-NOTstd} and \ref{sec-HSTstd}.
Fig.\ \ref{fig-apmag} show the comparison of the aperture magnitudes.
We find the mean differences to be 
$\Ic({\rm NOT})-\Ic({\rm HST}) = 0.066\pm 0.007$ and $0.017\pm 0.010$
for A665 and A2218, respectively.
Galaxies with aperture magnitude fainter than $19\mag$ were excluded
from this comparison in order to limit the errors introduced by
uncertainty in the sky subtraction.

\begin{figure}
\epsfxsize=8.8cm
\epsfbox{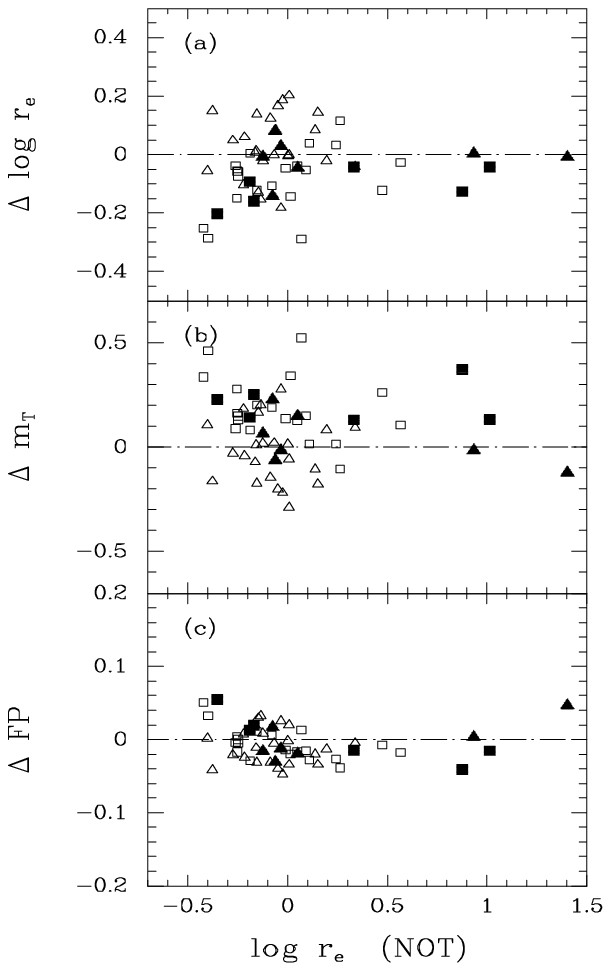}

\caption[ ]{Comparison of effective parameters in the I-band.
Boxes -- A665; triangles -- A2218.
Solid symbols -- galaxies for which we have spectroscopy.
The differences are calculated as ``NOT''--``HST''.
The formal errors derived from the fitting are of the size of the
points.
``FP'' is the combination $\log \re - 0.328\mue$ which enters the FP.
Only galaxies with total magnitudes brighter than $19\mag$ are shown.
The HST photometry shown on the figure has {\sl not} been offset to 
consistency with the ground-based photometry.
\label{fig-eff} }
\end{figure}

\begin{figure}
\epsfxsize=8.8cm
\epsfbox{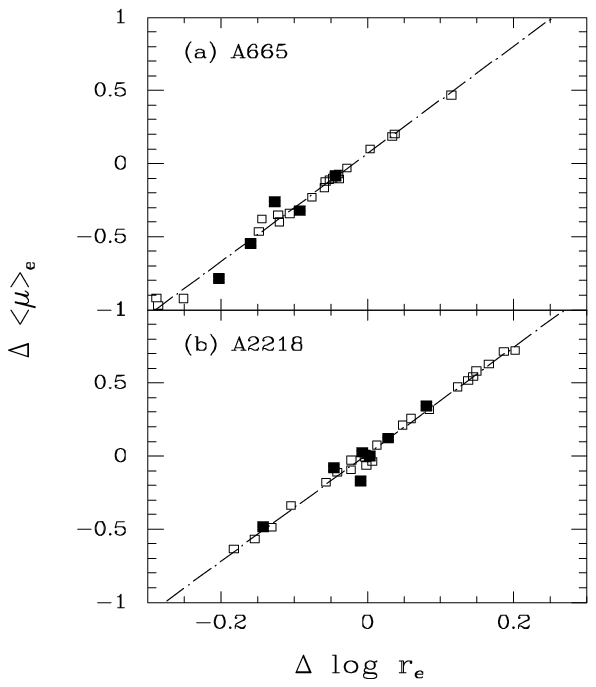}

\caption[ ]{The difference in $\log \re$ versus the difference 
in $\mue$.  The differences are calculated as ``NOT''-``HST''.
Only galaxies with total magnitudes brighter than $19\mag$ are shown.
Solid symbols -- galaxies for which we have spectroscopy.
The dashed lines show least squares fits to the data.
The offset in $\Delta \mue$ can be used to derive the offset between 
the HST photometry and the ground-based photometry, see text.
\label{fig-lremue} }
\end{figure}

\begin{table}
\caption[]{Comparison of effective parameters \label{tab-comp} }
\begin{tabular}{lrr}
Parameter  & mean difference   &   rms \\ \hline
$\log \re$ & $-0.033\pm 0.015$ & 0.11 \\
$\mue$     & $-0.119\pm 0.053$ & 0.41 \\
$\mT$      &  $0.044\pm 0.021$ & 0.16 \\
FP$^a$     &  $0.006\pm 0.003$ & 0.026 \\ \hline
\end{tabular}

Notes.  $^a$ the combination $\log \re - 0.328 \mue$.
The differences are calculated as ``NOT''--``HST''.
Only galaxies with total magnitudes brighter than $19\mag$ were included
in the comparisons. The two clusters are combined in this table,
and the HST photometry has been offset to consistency with the 
ground-based photometry, see text.
\end{table}

The effective parameters derived from the NOT images and the HST images
were compared, see Fig.\ \ref{fig-eff} and Table \ref{tab-comp}.
Only the galaxies with total magnitudes brighter than $19\mag$ in 
the I-band were included in this comparison, a total of 56 galaxies.
For the fainter galaxies the individual parameters ($\log \re$, $\mue$
and $\mT$) become highly uncertain due to the small size of the 
galaxies compared to the FWHM of the PSF for the ground-based data,
and due to rather low signal-to-noise for these faint galaxies in the
HST observations of A665. The combination ``FP''=$\log \re - 0.328\mue$,
which enters the FP, was also compared. This combination is well 
determined even for the fainter galaxies, but since none of the galaxies
for which we have spectroscopy are fainter than $19\mag$ we restrict
the comparisons to galaxies brighter than this limit.
The rms scatter for the comparison of the combination ``FP''
is similar to typical values for comparisons of data for nearby clusters
(e.g., J{\o}rgensen et al.\ 1995a). If the uncertainty is equally
distributed on two data sets, the typical uncertainty on the
combination ``FP'' is 0.018.
The small mean difference in the combination ``FP'' between the NOT data and
the HST data is equivalent to a difference in the FP zero point of 
only 1.4\%.  The comparisons of the parameters 
$\log \re$, $\mue$ and $\mT$ have slightly larger scatter than usually 
found for comparisons between effective parameters for nearby galaxies
derived by different authors.

The errors on $\log \re$ and $\mue$ are highly correlated
(e.g., J\o rgensen et al.\ 1995a). We use this fact to provide another
check of the magnitude zero points of the HST photometry.
Fig.\ \ref{fig-lremue} shows the difference in $\log \re$ versus the 
difference in $\mue$. The differences were calculated as
``NOT''-``HST''. 
The dashed lines on the panels show least squares fits to the data.
If the zero point for the HST photometry was in perfect agreement with 
the ground-based zero point, then the dashed lines would go through 
(0,0). The offset in $\Delta \mue$ is equal to the difference in the
zero points. 
We find $\Ic({\rm NOT})-\Ic({\rm HST}) = 0.063\pm 0.011$ and
$0.015 \pm 0.008$ for A665 and A2218, respectively.
This is in agreement with the determinations of the zero point 
differences based on aperture magnitudes.

Offsets of a few hundredths of a magnitude are common between HST 
photometry and ground-based photometry (e.g., Ellis et al.\ 1997;
Hill et al.\ 1998).
The offset we find for the A665 photometry is slightly larger than
expected. However, in order to ensure the consistency of the data for
A665 and A2218 we have chosen
to calibrate the HST photometry to the ground-based $\Ic$.
We have applied the following offsets:
$\Ic = \Ic({\rm HST}) + 0.065$ for A665 and
$\Ic = \Ic({\rm HST}) + 0.016$ for A2218.

We have no reliable way of checking the zero point for the V magnitudes
based on the HST data, because of the poor seeing of our ground-based 
V-images. We have therefore not applied any zero point correction 
to the colors $(V-\Ic )$ derived from the HST data.

\refstepcounter{section}

\begin{figure}
\epsfxsize=8.8cm
\epsfbox{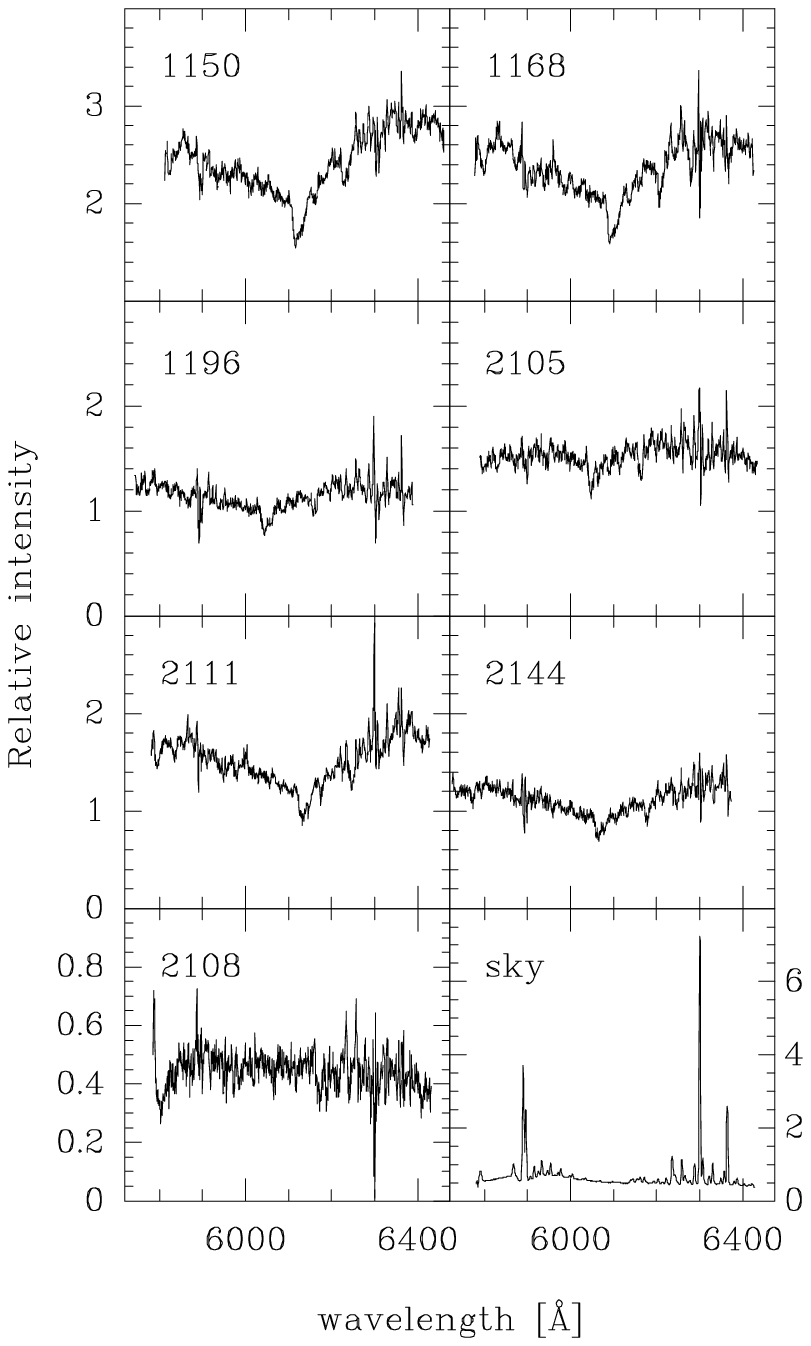}

\caption[]{Spectra of the observed galaxies in A665.
The spectra are not flux calibrated. \\
\label{fig-spA665}
\vspace*{-0.3cm} }
\end{figure}

\addtocounter{section}{-1}
\section{Spectroscopy}

\refstepcounter{figure}

\begin{figure*}
\epsfxsize=17.5cm
\epsfbox{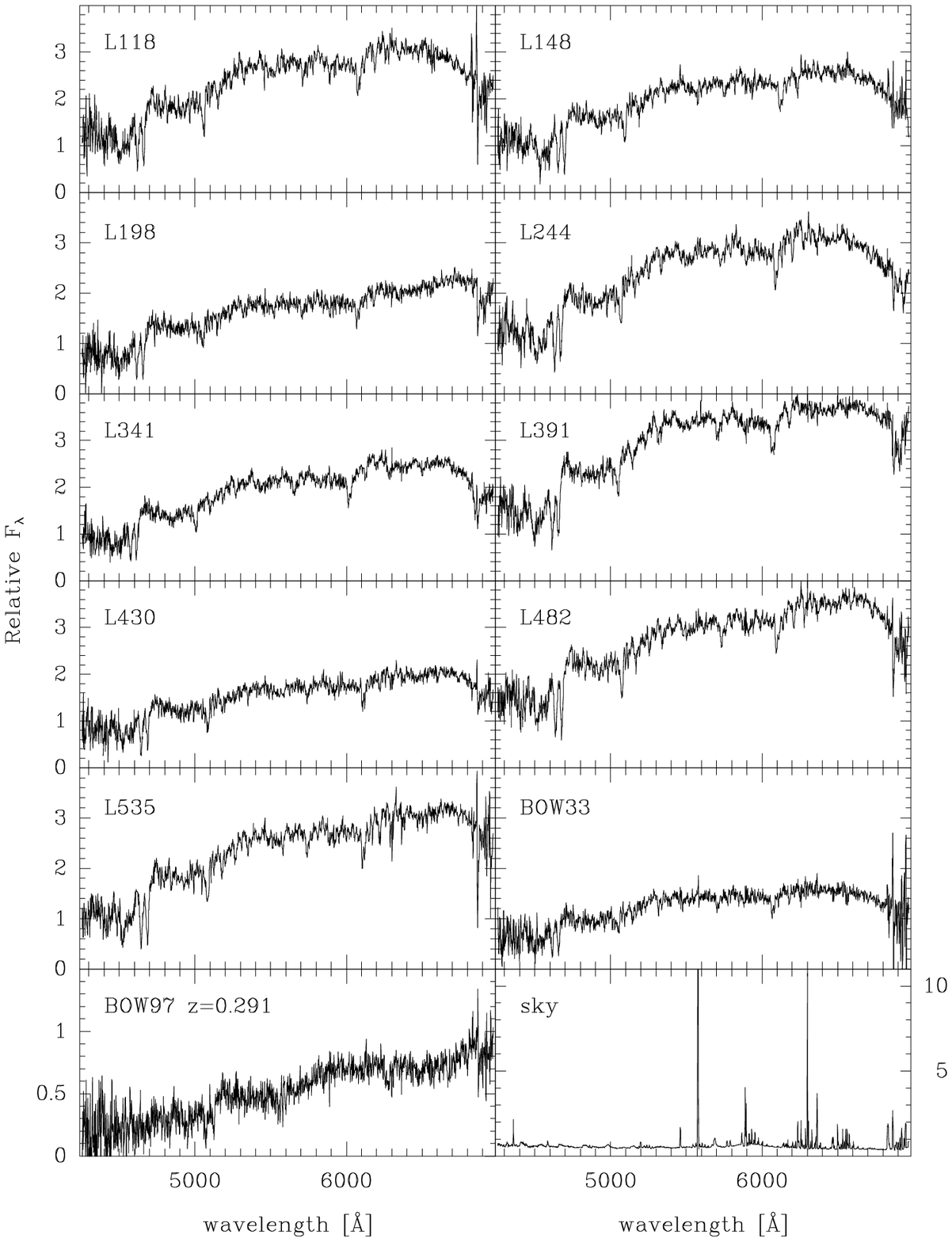}

\vspace*{-0.3cm}
\caption[]{Spectra of the observed galaxies in A2218.
The spectra are calibrated to a relative flux scale.
The panels are labeled with galaxy numbers from Le Borgne et al.\ (1992)
or Butcher et al.\ (1983).
BOW97 is a background galaxy at z=0.291.
\label{fig-spA2218} }
\end{figure*}

The spectroscopic observations were made through multi-slit masks.
One field was observed in each cluster.
The final spectra are shown in Figs.\ \ref{fig-spA665} and
\ref{fig-spA2218}.
The reductions of the A2218 spectra are described in the following.
The methods for the reductions of the A665 spectra were very similar.
We refer to Franx (1993ab) for additional information regarding
the A665 spectra.

\subsection{A2218}

The spectra were corrected for the bias and the dark by standard
methods. The bias subtraction was based on the level in the overscan
region as well as individual bias frames.

The flat fields show strong variations in the fringing depending on the 
zenith distance of the telescope due to the flexure of the instrument 
Therefore, flat fields obtained in day time did not give sufficiently
good flat field correction.
Instead we use flat fields obtained during the night.
These flat fields were obtained with the internal quartz lamp in the
instrument. Two flat fields bracket each science exposure.
The flat fields obtained in day time were combined in the sets they
were taken cleaned for cosmic-ray-events (CR).
Then all the flat fields were normalized
by fitting a high order spline in the direction of the dispersion.
In order to clean the night time flat fields for CRs
they were each compared to the 
the day time flat field, which best resembled night time flat field.
The final flat fields were derived as the average of the two flat fields,
which bracket the science exposure.
The typical pixel-to-pixel noise in the final flat fields was 0.6\%.
Areas of the flat fields that had a pixel-to-pixel noise larger than
1.5\% were set to one, and no flat field correction was performed.
This also ensures that no artificial variations are introduced at the 
edges of the slit-lets.

The individual exposures of A2218 were cleaned for CRs as follows.
First, the median image was derived.
Then difference between each individual image and the median image
was calculated. This image naturally has strong residuals from the
sky lines and the galaxies.
The residuals from the galaxies were fitted with a smooth function
in the direction of the dispersion, and the fit was subtracted.
The residuals from the sky lines were fitted with a constant across
each slit-let, and the fit was subtracted.
The resulting difference image contains noise and signal from 
CRs, plus some residuals from the strongest sky lines.
Pixels for which the deviations were below 7 times the local noise
or which had residuals from the sky lines were set to zero.
This resulted in a mask image for each of the science exposures. 
The mask image was subtracted from its corresponding science exposure.
The net-effect is that pixels affected by the CRs
get substituted with the median value, variations due to the shifts of
the galaxy spectra and the sky lines taken into account.

The spectra were corrected for the slit function based on sky flat fields.
The derived slit function was shifted up to 0.4 pixel in the spatial
direction in order to take into account the shifts in the slit-lets
due to the flexure.

The wavelength calibration was done for each slit-let individually.
The main wavelength calibration was established from high signal-to-noise
He-Ne-Ar lamp spectra taken in day time.
The dispersion solution was then shifted slightly based on He-Ne-Ar lamp 
spectra taken immediately before and after each science exposure.
The typical rms of the wavelength calibration is 0.18\AA .

The spectra were corrected for the S-distortion by tracing the position
of the spectrum of the BCG.
Because the signal in the individual exposures is too low to establish the 
S-distortion for the fainter galaxies the same S-distortion was used 
to correct all the galaxies. 
After the correction was applied, we verified that the same correction
could be use for all the galaxies by deriving the differences 
between the mean center of the the spectra in the wavelength
intervals 5295--5445\AA\ and 6040--6130{\AA}.
The BCG has a difference of 0.02 pixels, while the rest of the
galaxies show differences of less than 0.2 pixels. The aperture
used for the velocity dispersion measurements is 5 pixels.
We therefore conclude that the adopted correction for the S-distortion
does not affect the velocity dispersion measurements.

The wavelength calibration and the correction for the S-distortion
were applied simultaneously resulting in individual science exposures
that are all on the same wavelength scale and with the galaxies centered
on the same pixels.
The spectra were sampled on a logarithmic wavelength scale.
The individual images were then averaged with scaling according to 
their exposure time.
The individual expsure times were $11\times 1$h and one 
exposure of 48 minutes.
The average image was at this stage cut into sub-images,
each sub-image contains the spectrum from one slit-let.
The sub-images are then treated as individual long-slit spectra.

The sky background was subtracted from each sub-image based on the
outer $3\farcs 5$-$10\farcs 35$.
When possible the sky spectra were extracted from both sides of the 
galaxy spectra.
In some cases residuals were left from the stronger sky lines.
The sky subtraction of these lines was improved by fitting a linear
function to the residuals in the spatial direction, and then subtracting
the fit.
The residuals from the sky line at 5577\AA\ were interpolated across.

The spectra were calibrated to a relative flux scale based on exposures
of the flux standard star HD192281.
The flux calibrated spectra are in this paper used for display
purposes only. The determination of the velocity dispersions does not
depend on the flux calibration.

\end{document}